\documentclass{article}

\usepackage{arxiv}
\usepackage[utf8]{inputenc}
\usepackage[T1]{fontenc}
\usepackage{hyperref}
\usepackage{url}
\usepackage{booktabs}
\usepackage{amsfonts}
\usepackage{microtype}
\usepackage{graphicx}
\usepackage{natbib}
\usepackage{caption}
\usepackage{multirow}
\usepackage{array}
\providecommand{\Description}[1]{}
\graphicspath{{figures/}}
\newcolumntype{L}[1]{>{\centering\arraybackslash}p{#1}}
\newcolumntype{C}[1]{>{\centering\arraybackslash}p{#1}}
\newcommand{\EvalTableStyle}[1]{%
  \scriptsize
  \captionsetup{skip=4pt}
  \renewcommand{\arraystretch}{1.18}
  \setlength{\tabcolsep}{#1}%
}
\begin{document}

\title{RepoOMP: Repository-Aware Hotspot OpenMP Parallelization via Dependency-Aware Context Reduction}

\author{
  Yongjie Qian\thanks{ORCID: \href{https://orcid.org/0009-0006-9283-2796}{0009-0006-9283-2796}} \\
  Institute of Software, Chinese Academy of Sciences \\
  Beijing, China \\
  \texttt{qianyongjie25@mails.ucas.ac.cn}
  \And
  Ke Gao \\
  Institute of Software, Chinese Academy of Sciences \\
  Beijing, China \\
  \texttt{gaoke@iscas.ac.cn}
  \And
  Zhibin Zhang \\
  Institute of Software, Chinese Academy of Sciences \\
  Beijing, China \\
  \texttt{zhangzhibin25@mails.ucas.ac.cn}
  \AND
  Shaohui Peng \\
  Institute of Software, Chinese Academy of Sciences \\
  Beijing, China \\
  \texttt{pengshaohui@iscas.ac.cn}
  \And
  Ling Li \\
  Institute of Software, Chinese Academy of Sciences \\
  Beijing, China \\
  \texttt{liling@iscas.ac.cn}
}

\maketitle

\begin{abstract}
OpenMP parallelization of hotspots in mature repositories remains difficult because loop safety and optimization payoff often depend on non-local evidence. Rule-based tools under-parallelize when legality is not locally provable, while agent-based approaches become unstable when retrieval misses decisive dependencies or includes irrelevant code. We present RepoOMP, a hybrid framework that recovers parallelization-relevant evidence before generation. RepoOMP builds a Multi-granularity Attributes Performance graph (MAP), routes hotspots between deterministic rules and an LLM agent, and constructs a Structured Transformation Context (STC) that exposes dependency facts without flooding the model with unrelated repository text. We evaluate RepoOMP on 951 profiled hotspots from NPB, BOTS, FFmpeg, NCNN, and GROMACS. Under compilation, workload-specific checks, and positive speedup, 372 hotspots are accepted, including 330 real-world repository hotspots. RepoOMP achieves average speedups of $8.23\times$ on NPB and $8.96\times$ on BOTS. For the nine detailed real-world kernels used in matched-backbone and robustness analyses, RepoOMP reaches a cross-backbone mean of $5.25\times$, improves speedup by 18--28\%, and reduces agent-side token cost by 47--68\% relative to the unstructured Claude Code baseline. Across 330 accepted real-world hotspots, median speedup is $2.25\times$. Overall, RepoOMP provides an evidence-guided workflow for hotspot parallelization in repository settings. The open-source repository is available at \url{https://github.com/Qlalq/RepoOMP_Simplified}.
\end{abstract}

\keywords{repository-aware hotspot parallelization \and OpenMP \and program analysis \and large language models \and software performance \and hybrid optimization}

\section{Introduction}

OpenMP remains a practical mechanism for exploiting multicore CPUs in scientific and systems software\cite{openmp_ten}, and it continues to play an important role in shared-memory optimization for production software\cite{openmp_role}. Yet inserting correct and profitable directives into mature repositories is still labor intensive. In real repositories, whether a loop is safe and worthwhile to parallelize often depends on side effects hidden in helper functions, file-scope variables, and transitive callees distributed across multiple files. The challenge is to recover the non-local dependency evidence that determines legality and payoff. Hotspot-oriented parallelization in repository contexts is therefore a software engineering problem centered on evidence recovery. Classical compiler-based techniques are effective when the relevant dependence evidence stays close to the target loop, but they become less effective once the decisive facts escape the local region\cite{pluto}.

A running example illustrates this gap. Suppose a top-level loop iterates over a collection of tasks and invokes \texttt{compute(item)} for each element. Inside \texttt{compute}, a helper function \texttt{update\_stats()} writes to a file-scope array that records global execution statistics. From the local body of the outer loop, the iterations appear independent: each iteration processes a different item, and the shared write occurs several calls away. A rule-based tool that cannot determine whether \texttt{update\_stats()} may affect shared state will usually reject the transformation to avoid races, reflecting the conservative behavior of dependence-driven parallelizers when aliasing and transitive side effects become difficult to discharge statically\cite{cetus}. An LLM agent given only the local loop may make the opposite mistake and insert \texttt{\#pragma omp parallel for}, overlooking the hidden write and producing a wrong answer. Recent repository-level agent studies report the same pattern when context retrieval is incomplete\cite{autocoderover}. One failure loses a valid opportunity, while the other accepts an unsafe transformation.

\begin{figure}[t]
    \centering
    \includegraphics[width=0.90\linewidth]{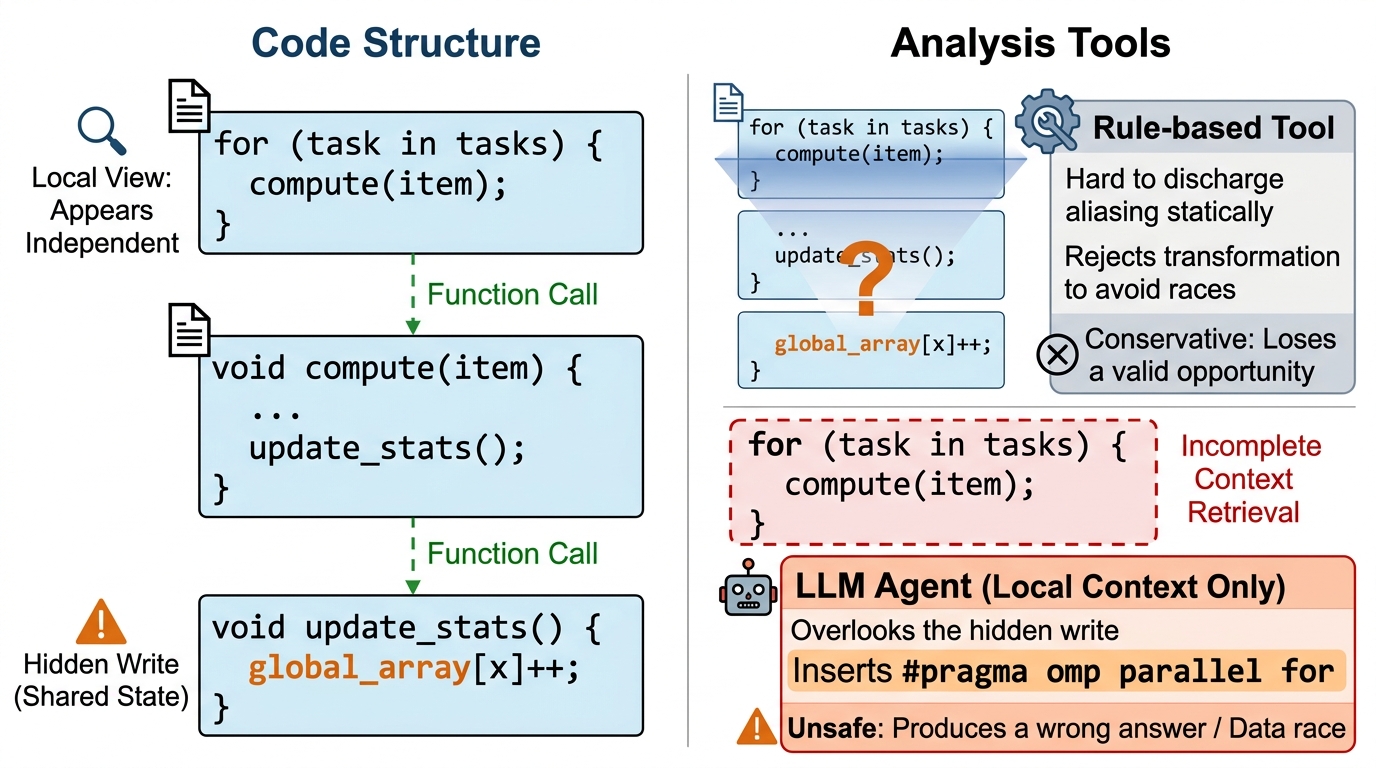}
    \caption{Running example motivating repository-context evidence recovery. A locally independent-looking loop reaches a hidden shared write through \texttt{compute(item)} $\rightarrow$ \texttt{update\_stats()}; rule-based tools may reject it conservatively, while an LLM given only local context may insert \texttt{\#pragma omp parallel for} and introduce a race.}
    \Description{A running example for repository-context auto-parallelization. On the left, the code structure shows a loop over tasks calling compute(item), which then calls update_stats(), where a hidden shared write increments global_array[x]. On the right, the figure contrasts two analysis outcomes. A rule-based tool rejects the transformation conservatively because it cannot safely discharge aliasing and shared-state effects. An LLM agent with incomplete local context overlooks the hidden write, inserts parallel-for, and becomes unsafe.}
    \label{fig:intro_running_example}
\end{figure}

Figure~\ref{fig:intro_running_example} makes this failure boundary explicit. The issue is whether the system can recover the transitive dependency evidence behind the local snippet. Existing tools often expose an interface mismatch for repository-scale hotspot optimization. Classical automatic parallelizers and source-to-source tools typically require a concrete file, loop region, or compilation unit as the optimization target; a repository maintainer, by contrast, often starts from an executable workload command and wants the system to discover which of many profiled files should be changed, which loops are legally parallelizable, and which changes preserve the command's observable behavior. This difference is illustrated by the command-level workloads in Table~\ref{tab:benchmarks_split}. Without an explicit repository-level evidence-recovery stage, a tool that accepts only a local optimization target cannot by itself map a workload command to the relevant source files and transformation sites, and exhaustively searching the resulting file--loop--context space would be too large to remain safe without strong pruning. Rule-based methods are reliable when the relevant dependencies remain visible to static checks, but they lose recall once pointer aliasing, indirect calls, or cross-function side effects obscure safety\cite{quinlan2011rose}. Their practical limitation is therefore an implicit local-provability assumption: they attempt parallelization only when legality can be discharged from a bounded static view. In repository contexts, however, the evidence for hotspot parallelization is distributed across multiple files and functions, making local proofs insufficient for many profitable candidates. Practical rule-based tools therefore restrict themselves to structurally regular cases with explicit dependence evidence and leave aside candidates whose legality would require cross-file reasoning over a very large search space. This preserves soundness, but it sacrifices profitable opportunities whose safety depends on non-local evidence. Engineering-oriented auto-parallelization studies report similar limitations when legality depends on evidence dispersed beyond the immediate loop body\cite{omp_engineer}.

Agent-based methods fail for the opposite reason. They can reason about nontrivial code and perform larger rewrites, but their success depends on context construction. If the prompt omits a distant dependency, the agent becomes unsafe. If it includes too much surrounding repository code, reasoning degrades, verification loops lengthen, and token cost rises sharply. Their core weakness is a retrieval assumption compounded by a practical modeling bias toward locally plausible pragma insertion: repository search optimized for functional relevance does not reliably recover the helper functions, file-scope variables, and transitive dependencies that determine parallel safety and performance payoff, while code agents often treat loops that look locally independent as attractive candidates for \texttt{\#pragma omp parallel for}. This heuristic can work on regular kernels, but it breaks in large repositories because the decisive evidence is often hidden behind project-specific helper functions, transitive side effects, and shared-state conventions that are not inferable from local syntax alone. Our own tables are consistent with this failure pattern: the unstructured agent baselines incur repeated BF/WA outcomes on repository workloads even when they sometimes find strong speedups on easier kernels. Recent repository-level generation studies report the same broader pattern: quality depends less on raw context scale than on whether tools recover the dependency evidence that generation actually requires\cite{toolgen_tosem}. Rule-based tools tend to under-parallelize, whereas agent-based systems tend to over-parallelize. The main bottleneck is recovering the evidence that determines whether a transformation is safe and worthwhile.

RepoOMP starts from a simple observation: hotspot-oriented parallelization in repository contexts fails when dependency structure is represented only as raw source text and retrieved only by functional similarity. The system must recover a repository-level view of data flow, performance hotspots, and scope boundaries, and then compress that evidence into a Structured Transformation Context (STC) for each optimization target. This need for structured context becomes more important as OpenMP programming itself has evolved toward richer constructs and more demanding optimization choices\cite{omp_80}. Recent repository-reasoning work points in the same direction: graph-backed retrieval improves large-code generation when the model is given structured evidence instead of undifferentiated repository text\cite{codexgraph}. In this setting, the LLM serves as a bounded transformation engine.

Based on this observation, we present RepoOMP, a hybrid framework for hotspot-oriented OpenMP parallelization in repository contexts. RepoOMP builds a Multi-granularity Attributes Performance graph (MAP), a repository abstraction over hotspot locations, call and dependency relations, and shared-state access summaries relevant to parallelization. MAP then drives a Rule-Agent Router, which selects between deterministic transformation, agent-based transformation, or conservative handling based on propagated dependency-risk cues. Candidates with clear dependency structure are sent to a deterministic rule engine, which preserves the low-risk advantage of explicit proofs without paying unnecessary model cost. Candidates whose main challenge is semantic restructuring under manageable dependency uncertainty are sent to an LLM agent together with a MAP-derived Structured Transformation Context (STC), the per-target condensed evidence package derived from MAP. Methodologically, RepoOMP is a repository-level evidence-recovery workflow built from MAP, bounded STC construction, and failure-mode-aware routing. This design follows a broader lesson from hybrid software engineering systems: program analysis should recover task-critical structure before the model attempts generation or repair\cite{hybrid_apr_tosem}. In the running example, the resulting context includes the target loop, the file-scope state it touches, and the helper summaries through which the shared write is introduced. RepoOMP addresses the task through evidence recovery, failure-mode-aware routing, and bounded transformation.

RepoOMP also closes the loop with explicit verification. Each generated transformation is evaluated through compilation, workload-specific executable checks, and performance measurement. Prior repository-level repair work similarly shows that executable validation is essential when models operate on nontrivial software artifacts, because textual plausibility alone does not guarantee buildability, semantic preservation, or operational usefulness\cite{morepair_tosem}. In this study, transformations are treated as accepted only when they compile, pass the reported workload-specific checks, and show positive performance under the specified configurations.

This paper makes the following contributions:
\begin{itemize}
    \item We identify two recurring failure modes in repository-scale OpenMP parallelization: conservative misses from local static proofs and unsafe transformations from incomplete agent context. This view motivates the routing design of RepoOMP.
    \item We introduce MAP, a repository abstraction for hotspot discovery, dependency-risk propagation, and STC construction.
    \item We design a Rule-Agent Router that sends structurally clear candidates to deterministic transformation and reserves LLM-based rewriting for candidates that need broader semantic reasoning.
    \item We evaluate RepoOMP on 951 profiled hotspots, including 330 accepted real-world repository hotspots. RepoOMP reaches average speedups of $8.23\times$ on NPB and $8.96\times$ on BOTS; for the nine detailed accepted real-world kernels used for matched-backbone and robustness analyses, it reaches $5.25\times$ and reduces agent-side token cost by 47--68\% relative to the unstructured Claude Code baseline used in this study.
\end{itemize}

\section{Related Work}

Prior work explains why repository-scale OpenMP generation remains difficult. Rule-based tools, specialized learned models, and LLM-based agents differ in accuracy and in how they assume optimization-relevant evidence can be accessed. They therefore fail at different evidence boundaries of the same task: judging whether a transformation is safe and worthwhile often requires dependency facts distributed across repository, file, and function scopes.

\subsection{Rule-based Tools}

Classical rule-based systems rely on static dependence analysis, symbolic reasoning, or polyhedral transformations to identify legal loop parallelism. Systems such as Pluto\cite{pluto}, Polly\cite{polly}, Cetus\cite{cetus}, Par4All\cite{par4all}, and AutoPar\cite{quinlan2011rose} are most effective when loop bounds, memory accesses, and dependence relations remain regular enough to be proven. Source-to-source engineering frameworks such as Clava-based auto-parallelization extend this line toward more practical transformation pipelines\cite{autopar_clava}. Dynamic and analysis-enhanced systems further widen coverage by introducing runtime evidence or speculative reasoning, as illustrated by DiscoPoP\cite{discopop} and Apollo\cite{apollo}.

This literature produces low-risk transformations when the relevant evidence is explicit, observable, or formally derivable. Its limitation appears when parallel safety depends on facts that are difficult to prove locally, such as transitive side effects, file-scope shared state, opaque aliasing, or irregular cross-function control flow. RepoOMP preserves the low-risk advantage of rule engines, but changes their role: they serve as the high-confidence branch after repository-level dependency summarization separates structurally clear candidates from cases that require richer reasoning.

\subsection{Specialized Pretrained Models}

A second line of work uses learned models to predict directives or optimization decisions directly from code. HPC-Coder-v2\cite{hpc-coder-v2}, PragFormer\cite{pragformer}, OMPar\cite{ompar}, AutoParLLM\cite{autoparllm}, and ConTraPh\cite{contraph} show that learned representations can capture useful parallelization patterns beyond hand-written heuristics, especially when the decision remains close to local pragma placement, structured kernels, or benchmark distributions seen during training. Recent TOSEM work in software engineering also suggests that learning becomes more robust when paired with auxiliary evidence recovery. RunTyper combines dynamic analysis with learned prediction to stabilize difficult type-inference decisions\cite{runtyper_tosem}, and MORepair shows that task-specific learning objectives can improve repository-level repair when plain next-token generation is too weak to reconstruct the needed constraints\cite{morepair_tosem}.

At repository scale, however, training locality and deployment locality diverge. Specialized models are often trained on local windows or controlled optimization benchmarks, while the decisive evidence for repository auto-parallelization may reside in another file, a transitive callee, or a shared-state summary that must be recovered before generation. ToolGen similarly shows that repository-level generation depends on restoring structured evidence rather than supplying more raw context\cite{toolgen_tosem}, and recent LLM-based program-analysis work shows that raw textual access is an unstable proxy for the facts needed by downstream reasoning\cite{llm_intro_3}. RepoOMP therefore first recovers optimization-relevant repository evidence with MAP, then asks the model to operate over that bounded abstraction.

\subsection{LLM-based Agents and Repository Tools}

Recent code agents and repository tools have substantially expanded the scope of automated programming systems. General-purpose coding agents such as Claude Code\cite{claudecode} and Codex\cite{codex_openai_2025} provide an important deployment backdrop, while research systems such as SWE-agent\cite{swe-agent}, OpenHands\cite{openhands}, and Devin\cite{devin} show that agents can navigate files, invoke tools, propose edits, and iterate against execution feedback. Industrial repository assistants point in a similar direction by emphasizing retrieval-augmented workflows over raw prompt expansion\cite{windsurf}. A parallel academic line shows that repository reasoning becomes more stable when retrieval, graph structure, planning, and program-analysis constraints are made explicit, as seen in AutoCodeRover\cite{autocoderover}, CodexGraph\cite{codexgraph}, CodePlan\cite{codeplan}, ToolGen\cite{toolgen_tosem}, and hybrid analysis-guided repair systems\cite{hybrid_apr_tosem}.

For repository-scale auto-parallelization, this literature reveals a different failure boundary from rule-based tools. The main issue is retrieval mismatch. Functional relevance and parallelization relevance diverge. An agent may retrieve code that explains what a feature does while still missing the helper function, file-scope variable, or transitive dependency that determines thread safety and performance payoff. Similar gaps between semantically useful context and transformation-critical context have also been observed in augmented code-completion settings\cite{augmentedcode}. Simply enlarging the context window does not resolve this reliably: small prompts can omit decisive evidence, whereas large prompts dilute reasoning and make iterative repair expensive. The underlying flaw is that generic repository retrieval is optimized for semantic usefulness instead of optimization-critical evidence recovery. RepoOMP adopts the broader lesson from repository agents and tool-use systems that structure should guide search, then applies that lesson to parallelization evidence. MAP, the Structured Transformation Context (STC), and the Rule-Agent Router recover the cross-layer facts that govern safe and profitable parallelization.

\begin{figure*}[t!]
    \centering
    \includegraphics[width=0.90\textwidth]{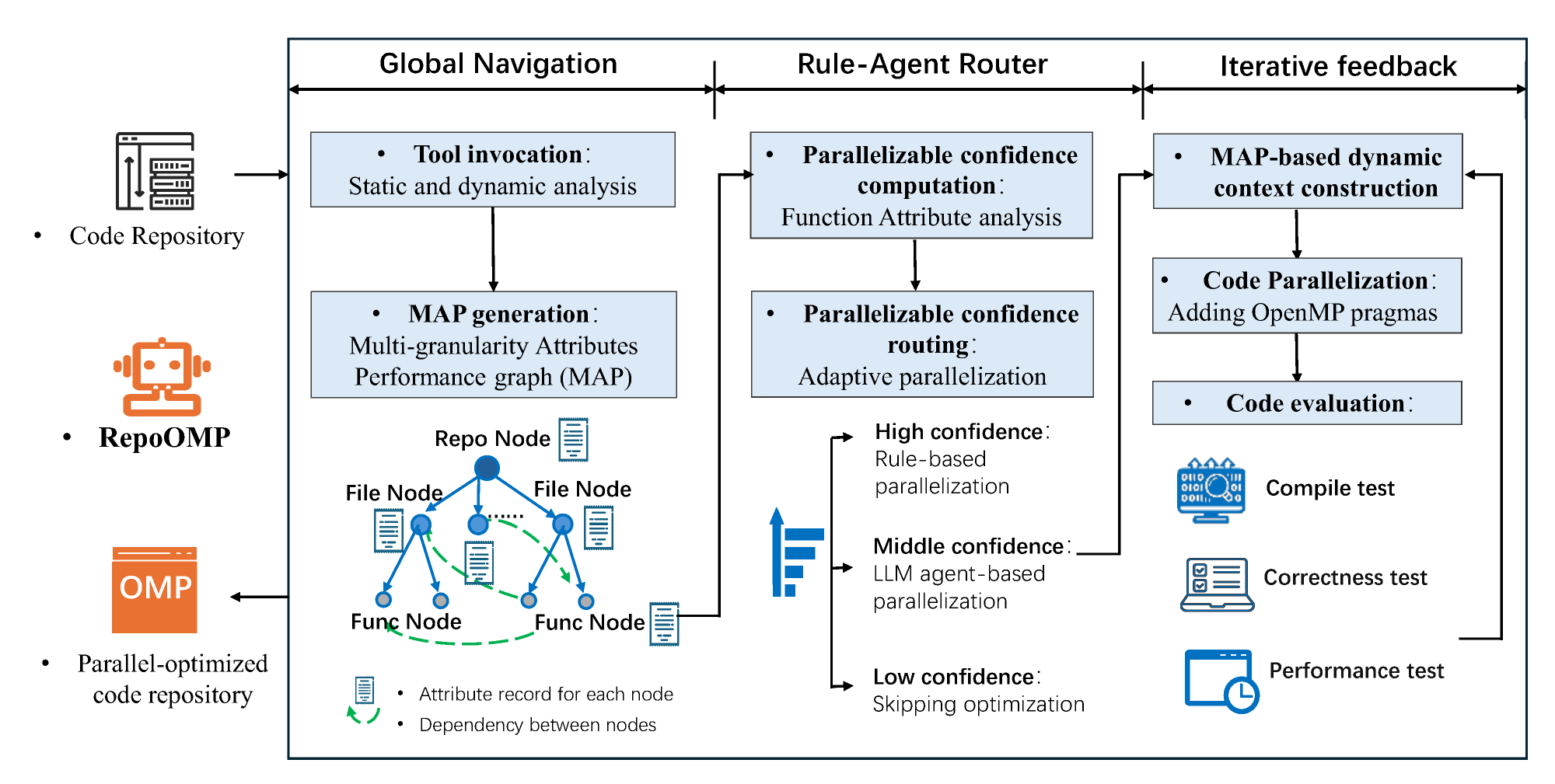}
    \caption{Overview of the RepoOMP framework. The workflow consists of three stages: Global Navigation constructs the Multi-granularity Attributes Performance graph (MAP) from the code repository; the Rule-Agent Router assesses parallelization confidence to triage tasks between rule-based and LLM-based parallelization engines; and Iterative Feedback validates correctness and performance to refine the optimized repository.}
    \Description{An overview diagram of RepoOMP with three connected stages. The first stage, Global Navigation, builds a multi-granularity graph over repository, file, and function levels and annotates nodes with dependency and runtime information. The second stage, the Rule-Agent Router, classifies candidate loops by parallelization confidence and dispatches them either to a deterministic rule engine or to an LLM-based agent with bounded context. The third stage, Iterative Feedback, validates candidate transformations through compilation, correctness checking, and performance measurement, then accepts or rolls back changes based on the results.}
    \label{fig:2}
\end{figure*}

\begin{table}[htbp]
    \centering
    \caption{Schema of MAP nodes and edges with representative fields}
    \label{tab:graph_schema}
    \scriptsize
    \renewcommand{\arraystretch}{1.18}
    \setlength{\tabcolsep}{3pt}
    \begin{tabular}{L{0.12\columnwidth}L{0.12\columnwidth}L{0.16\columnwidth}L{0.28\columnwidth}L{0.19\columnwidth}}
    \toprule
    \textbf{Scope} & \textbf{Type} & \textbf{Category} & \textbf{Key Fields} & \textbf{Role} \\
    \midrule
    \multirow{5}{*}{\textbf{Nodes}}
     & \textsc{Repo} & Global & \texttt{id}, \texttt{total\_time} & Repository-level runtime anchor \\
     \addlinespace 
     & \textsc{File} & Context & \texttt{id}, \texttt{static\_vars} & Exposes intra-file shared state \\
     \addlinespace
     & \multirow{3}{*}{\textsc{Func}} & Metrics & \texttt{total\_time}, \texttt{inclusive\_time}, \texttt{call\_count}, \texttt{avg\_time} & Ranks optimization hotspots \\
     & & Attributes & \texttt{start/end\_line}, \texttt{self\_status}, \texttt{global\_memory\_access} & Captures local code evidence \\
     & & Summary & \texttt{parallelism\_safety}, \texttt{blockers}, \texttt{global\_reads}, \texttt{global\_writes} & Summarizes safety and shared accesses \\
     \addlinespace

    \multirow{2}{*}{\textbf{Edges}}
     & Control & Flow & \texttt{calls}, \texttt{contains} & Links callees and hierarchy \\
     \addlinespace
     & Data & Dependency & \texttt{reads}, \texttt{writes}, \texttt{defines} & Tracks global data dependence \\
    \bottomrule
    \end{tabular}
\end{table}

\section{Method}

RepoOMP is built on the premise introduced in Section~1: hotspot-oriented parallelization in repository contexts should recover dependency evidence before deciding how code should be transformed. Figure~\ref{fig:2} summarizes the workflow: RepoOMP builds an optimization-relevant repository abstraction, routes each candidate according to its failure mode, and validates each generated transformation through compilation, workload checks, and timing. The methodological object of this section is the evidence-recovery pipeline itself: MAP exposes repository-level dependency structure, the router turns failure modes into explicit decisions, and STC construction converts that structure into a bounded transformation interface for the LLM.

\subsection{Problem Statement and Design Goals}

Given a compilable repository, RepoOMP aims to identify computation hotspots that may benefit from OpenMP parallelization and synthesize source-level transformations that preserve workload-level validity while improving performance. Loop safety is often determined by non-local information. In the running example, the outer loop appears independent until repository-level analysis reveals that one helper writes shared file-scope state. This motivates three design goals.

First, the system should recover dependencies across repository, file, and function boundaries instead of relying on isolated loop bodies. Second, it should distinguish candidates clear enough for deterministic rules from candidates that require semantic rewriting. Third, when an LLM is used, the prompt should include the dependency facts needed for the decision while excluding unrelated code. These goals define RepoOMP as evidence recovery, selective routing, and bounded transformation.

\subsection{MAP: Dependency- and Performance-Aware Repository Abstraction}

To satisfy these goals, RepoOMP builds the Multi-granularity Attributes Performance graph (MAP), a directed property graph $G=(V,E)$ that summarizes optimization-relevant repository structure. MAP contains three node granularities. A repository node stores repository-wide optimization metadata. File nodes capture file-scope information, especially static variables whose visibility is larger than a single function but smaller than the whole repository. Function nodes serve as the main optimization units and record source locations, dependency summaries, and runtime attributes. Table~\ref{tab:graph_schema} lists the node and edge schema.

MAP is introduced because a function-level view is often insufficient even when the final transformation is local. In the running example, the outer loop can only be judged correctly if the system knows that one callee eventually writes a file-scope statistics array while another only updates thread-local state. MAP makes this distinction explicit through control-flow edges for call relations and data-dependency edges for reads, writes, and symbol definitions. We construct the graph using a compilation database generated by Bear and syntactic extraction from tree-sitter, then augment it with repository structure and source-level attributes relevant to OpenMP safety.

MAP differs from repository graphs designed for general code generation or navigation. Systems such as CodePlan focus on functional relevance, whereas RepoOMP needs optimization relevance. For this task, the crucial question is which transitive callees, globals, and side effects determine whether a loop is safe and profitable to parallelize. MAP is therefore specialized for dependency recovery and hotspot isolation.

\subsection{Confidence Propagation and Routing}

Once MAP is built, RepoOMP assigns each function a parallelization confidence category that approximates how likely a candidate is to fail under rule-based or agent-based processing. We begin with local evidence such as self-dependencies, file-scope writes, input/output operations, indirect memory accesses, and external calls. We then propagate this evidence bottom-up along the call graph so that callers inherit the risk introduced by their descendants.

This propagation is necessary because failure modes in repository contexts are transitive. A caller may appear simple while depending on a low-confidence callee that mutates shared state. In the running example, the top-level loop is not truly high confidence because one nested helper eventually writes a shared statistics array. Propagation turns that hidden dependency into an explicit routing signal.

The Rule-Agent Router uses these propagated summaries to choose among three branches. High-confidence candidates are handled by the rule engine; these cases typically include regular loop patterns or reductions whose safety remains clear after dependency summarization. Middle-confidence candidates are routed to the LLM agent; these cases often involve restructurable logic, irregular control flow, or incomplete local evidence, but their dependency footprint can still be summarized into a bounded context. Low-confidence candidates are treated as the blocker-dominated tail in the hotspot funnel: under the reported workflow, they are not expected to become high-yield optimization cases, although the outcome table is anchored by the final observed build, check, rollback, and acceptance results instead of branch labels alone.

In the current implementation, the router is a deterministic engineering heuristic, not a learned probabilistic classifier. High-confidence routing is used when propagated summaries preserve a regular parallel structure without blocker attributes; low-confidence routing is used when propagated summaries retain unresolved blockers such as shared-state writes, I/O, indirect accesses, or serial-control constraints; and the remaining candidates are treated as middle confidence and passed to the agent. We present this policy as an explicit bounded-decision heuristic for repository optimization.

The router operationalizes the two dominant error modes discussed earlier. Rules are effective at avoiding unsafe transformations but tend to miss opportunities. Agents can recover some of those opportunities, but only when the dependency context is trustworthy and sufficiently bounded. The router turns this asymmetry into an explicit decision policy instead of assuming that one solver should handle all candidates.

Operationally, the current heuristic records for each candidate a local attribute tuple consisting of loop regularity, reduction-like structure, file-scope or global reads and writes, I/O calls, serial-control markers, indirect memory accesses, external calls, and the directly reachable callee set. Propagation then computes the union of blocker attributes over the transitive callees and marks a shared-state hazard whenever any reachable summary contains a file-scope or global write that is not discharged as a reduction-like update. The router applies a bounded three-way rule to this propagated tuple: a candidate is \emph{high confidence} if it retains a regular loop or reduction structure and the propagated blocker set is empty; \emph{low confidence} if the propagated blocker set contains a shared-state hazard, I/O, serial control, or unresolved indirect or external effects; and \emph{middle confidence} otherwise. Each routing decision is therefore inspectable from three stored artifacts: the local attributes, the propagated blocker set, and the final branch label.

Table~\ref{tab:routing_signals} summarizes representative predicates and actions for the three branches.

\begin{table}[t]
\centering
\caption{Representative routing predicates and downstream actions in RepoOMP.}
\label{tab:routing_signals}
\scriptsize
\setlength{\tabcolsep}{3.5pt}
\renewcommand{\arraystretch}{1.1}
\begin{tabular}{C{0.18\columnwidth}C{0.27\columnwidth}C{0.18\columnwidth}C{0.24\columnwidth}}
\toprule
\textbf{Signal Class} & \textbf{MAP Predicate} & \textbf{Router Output} & \textbf{System Action} \\
\midrule
Regular parallel structure & No blocker in MAP and no shared-state write & High confidence & Rule engine only \\
\addlinespace
Bounded dependency uncertainty & Incomplete local evidence, but relevant callees, \texttt{global\_reads}, and \texttt{global\_writes} remain bounded in MAP & Middle confidence & Build Structured Transformation Context (STC), then call LLM \\
\addlinespace
Indirect access with bounded summaries & Indirect access or external calls remain, but their summaries stay bounded in MAP & Middle confidence & Build Structured Transformation Context (STC) with callee summaries, then call LLM \\
\addlinespace
I/O or serial control dependence & I/O or serial-control blocker appears in MAP & Low confidence & Conservative handling; keep serial code unless later evidence justifies change \\
\addlinespace
Shared-state hazard & Shared-state write or unresolved side effect appears in MAP & Low confidence & Skip transformation; record blocker summary \\
\bottomrule
\end{tabular}
\end{table}

\subsection{Structured Transformation Context Construction}

For candidates routed to the agent, RepoOMP constructs a Structured Transformation Context (STC) instead of sending either a single local function or a large raw repository slice. The STC combines three sources: a repository slice that provides the target code region to be transformed, MAP queries that recover file-level definitions and shared variables visible to that region, and LLM-generated summaries of the relevant MAP subgraph for transitive call-chain nodes whose effects matter for dependency reasoning.

In the running example, the repository slice contributes the outer loop that is the actual transformation target. MAP then contributes the definition of the shared statistics array and other file-scope state visible from that loop. Finally, an LLM summarizes the MAP-connected call-chain nodes into compact dependency-oriented descriptions indicating where shared writes or other blocker effects are introduced. The prompt therefore excludes semantically nearby but irrelevant files while preserving the specific evidence needed for thread-safety reasoning. This selective construction addresses both context failure modes: it avoids false positives caused by clipping away a hidden dependency, and it avoids reasoning degradation caused by flooding the model with irrelevant code.

Operationally, RepoOMP constructs STC in three steps. It first locates the candidate source span from the repository slice. It then queries MAP attributes such as \texttt{global\_reads}, \texttt{global\_writes}, symbol definitions, and blocker summaries to recover the visible shared-state boundary. Finally, it summarizes the relevant MAP neighborhood into short dependency-oriented descriptions for the transitive call-chain nodes that influence the decision. This separation of target code, graph-extracted dependency facts, and callee summaries gives the agent a stable reasoning interface while preserving the task-relevant evidence needed for transformation.

The Structured Transformation Context (STC) is serialized as five fields: the target source span to be transformed, the visible shared-state definitions at file scope, symbol definitions required to resolve the loop body, short summaries for the transitive callee nodes that remain on the MAP path to the target, and a constraint block stating blocker facts that must be preserved during transformation and validation. The agent prompt follows the same fixed template for all reported tasks: \emph{transform only the target span; use only the supplied dependency evidence; preserve buildability and the reported output checks; avoid introducing synchronization-free shared writes; emit no repository-wide rewrite beyond the bounded span}. This fixed interface keeps the reasoning scope aligned with the recovered dependency evidence and makes the agent path directly inspectable.

The main inspectable intermediates in this workflow are the compilation database, the MAP summaries, the router outcome for each candidate, the serialized STC, and the final validation result. Together, these artifacts provide a concrete trail from evidence recovery to transformation and acceptance, so that the reported outcomes can be traced to explicit intermediate representations rather than to unconstrained repository-scale prompting.

\subsection{Verification and Rollback Workflow}

After either the rule engine or the agent proposes an OpenMP transformation, RepoOMP executes a three-stage verification workflow. First, a compilation check filters out syntactic and build-system errors. Second, a workload-specific executable check compares program outputs against a golden baseline using application-specific validation, including MD5-based output matching where applicable. Third, a performance check measures whether the validated transformation yields meaningful speedup. For the reported real-world task-level results, the speedup and token values are arithmetic means over five repeated executions of the same generated implementation under the same workload input and reported 16-thread configuration.

These stages guard against different classes of failure. Compilation catches malformed directives and interface mismatches. Workload-specific executable checks catch observable semantic regressions under the tested configurations. Performance evaluation prevents the system from accepting transformations that are legal but unhelpful. If any stage fails, RepoOMP rolls the repository back to the previous valid state and records the candidate as unsuccessful. The current TSan audit is likewise reported under matched 16-thread command templates for the nine real-world workloads.

For reproducibility, the workflow has two replay levels. At the command level, Table~\ref{tab:benchmarks_split} gives the application command fragments and kernel arguments, while Table~\ref{tab:validation_oracle} specifies the input identity, oracle, tolerance, thread settings, and repeated-run policy used to decide acceptance. A complete artifact replay additionally requires the repository revision, build wrapper, compilation database, prepared workload assets (for example videos, model files, and \texttt{.tpr} inputs), profiling logs, serialized MAP, router labels, STC prompt records, generated patches, validation logs, timing logs, and token accounting for each candidate. In our implementation, these records are emitted as inspectable per-candidate artifacts so that an accepted result can be traced from workload command to hotspot selection, context construction, patch, validation, and timing. The submitted artifact is intended to package these records for the reported kernels, with large external assets referenced by checksum and setup script when redistribution is impractical. The command fragments in Table~\ref{tab:benchmarks_split} are therefore workload identifiers rather than a complete push-button artifact by themselves; the full replay package must include the wrappers and assets listed above.

This verification loop is essential because hotspot-oriented parallelization is a software engineering task with operational consequences. A plausible patch is insufficient. The transformation must compile in the target repository, preserve behavior under the tested workloads, and provide measurable benefit. Verification therefore closes the loop between dependency recovery, selective generation, and executable validation. In this study, a failed attempt denotes either a build failure or a mismatch under the task-specific executable checks listed in Table~\ref{tab:validation_oracle}, and an accepted transformation denotes a candidate that compiles, passes those checks, and delivers positive speedup under the tested configuration.

The evaluation therefore uses layered validation semantics rather than a single oracle. Compilation and workload-specific executable checks define acceptance, while the later ThreadSanitizer audit provides an additional workload-level dynamic concurrency signal on the reported real-world workloads. Table~\ref{tab:validation_oracle} makes the acceptance checks explicit, and Table~\ref{tab:tsan_summary} reports the supplementary audit under the matched 16-thread command templates.

\begin{table}[t]
\centering
\caption{Validation oracle per workload. \textit{Oracle} denotes the correctness reference, \textit{Tol.} denotes the acceptable numerical tolerance, and \textit{Check} classifies the validation type. Functional checks require exact output match; numerical checks tolerate bounded deviation. For the reported real-world task-level results, the final column records the workload identity together with the bounded check/timing protocol: executable checks are exercised across the reported thread settings, while the runtime cells in Tables~\ref{tab:large_repo_results} and \ref{tab:large_repo_results_cont} are five-run arithmetic means at 16 threads.}
\label{tab:validation_oracle}
\scriptsize
\renewcommand{\arraystretch}{1.15}
\setlength{\tabcolsep}{2.5pt}
\resizebox{\columnwidth}{!}{%
\begin{tabular}{ccccc}
\toprule
Workload & Oracle & Tol. & Check & Workload / Runs$\times$Threads \\
\midrule
NPB (all) & Reference output files & Exact & Functional & Class A/B sizes \\
BOTS (all) & Task-specific reference & Exact & Functional & Default problem size \\
FFmpeg-MPEG4 & Frame-level MD5 hash & Exact & Functional & Synthetic 1080p video; chk $1/16/32/64$T, time $5\times@16$T \\
FFmpeg-NLMeans & PSNR threshold & $>40$dB & Numerical & Noisy 1080p frame; chk $1/16/32/64$T, time $5\times@16$T \\
FFmpeg-Deshake & Motion-vector check & Bounded & Functional & Handheld 1080p clip; chk $1/16/32/64$T, time $5\times@16$T \\
NCNN-MobNet & Top-5 accuracy & $<0.1\%$ drop & Numerical & ImageNet subset; chk $1/16/32/64$T, time $5\times@16$T \\
NCNN-ShuffNet & Top-5 accuracy & $<0.1\%$ drop & Numerical & ImageNet subset; chk $1/16/32/64$T, time $5\times@16$T \\
NCNN-ResNet & Top-5 accuracy & $<0.1\%$ drop & Numerical & ImageNet subset; chk $1/16/32/64$T, time $5\times@16$T \\
GROMACS-Nonbd & Energy drift & $<10^{-5}$ & Numerical & 5000-step trajectory; chk $1/16/32/64$T, time $5\times@16$T \\
GROMACS-PME & Energy drift & $<10^{-5}$ & Numerical & 5000-step trajectory; chk $1/16/32/64$T, time $5\times@16$T \\
GROMACS-LINCS & Constraint residual & $<10^{-4}$ & Numerical & 5000-step trajectory; chk $1/16/32/64$T, time $5\times@16$T \\
\bottomrule
\end{tabular}
}
\end{table}

Across the nine reported real-world kernels, this \emph{Runs$\times$Threads} protocol yields a clear summary: eight kernels pass all executable checks on all five repeated runs at 1, 16, 32, and 64 threads, while FFmpeg-Deshake passes all five runs at 1/16/32 threads and four of five runs at 64 threads. We therefore report task-level performance at 16 threads as five-run means and report the broader thread-sweep checks separately as cross-thread stability evidence for the same workloads.

\begin{table*}[t]
\centering
\begin{minipage}[t]{0.38\textwidth}
\centering
\caption{Statistics of the evaluated datasets. The MAP Scale reflects the complexity of the performance graph constructed by RepoOMP.}
\label{tab:datasets}
\small
\resizebox{0.98\linewidth}{!}{%
\begin{tabular}{cccccc}
\toprule
\multirow{2}{*}{Metric} & \multicolumn{2}{c}{Micro-Benchmarks} & \multicolumn{3}{c}{Real-world Apps} \\
\cmidrule(lr){2-3} \cmidrule(l){4-6}
 & NPB & BOTS & NCNN & FFmpeg & GROMACS \\
\midrule
Files & 14 & 106 & 1541 & 3245 & 2778 \\
Functions & 36 & 323 & 2539 & 18952 & 6959 \\
LOC (k) & 2.58 & 24.04 & 712.76 & 1160.63 & 811.15 \\
Kernels & 8 & 8 & 3 & 3 & 3 \\
\midrule
\multicolumn{6}{l}{MAP Scale} \\
Nodes (k) & 0.05 & 0.43 & 4.08 & 22.20 & 9.74 \\
Edges (k) & 0.08 & 1.23 & 22.95 & 90.29 & 23.36 \\
\bottomrule
\end{tabular}%
}
\end{minipage}\hfill
\begin{minipage}[t]{0.58\textwidth}
\centering
\caption{Execution Configurations. Shared command prefixes are factored out by application, while the kernel-specific arguments are listed separately.}
\label{tab:benchmarks_split}
\vspace{\baselineskip}
\scriptsize
\renewcommand{\arraystretch}{0.95}
\setlength{\tabcolsep}{2.5pt}
\resizebox{0.98\linewidth}{!}{%
\begin{tabular}{@{}cccc@{}}
\toprule
\textbf{Application} & \textbf{Prefix} & \textbf{Kernel} & \textbf{Args} \\
\midrule

\multirow{3}{*}{\textbf{FFmpeg}}
 & \multirow{3}{*}{\texttt{./ffmpeg -threads 1 -i input.mp4}} & MPEG4   & \texttt{-c:v mpeg4 -b:v 2M -f null -} \\
 &  & NLMeans & \texttt{-vf nlmeans -f null -} \\
 &  & Deshake & \texttt{-vf deshake -f null -} \\
\midrule

\multirow{3}{*}{\textbf{NCNN}}
 & \multirow{3}{*}{\texttt{./benchncnn 10 16 0 0 1}} & MobileNet  & \texttt{param=mobilenet\_v3 shape=[224,224,3]} \\
 &  & ShuffleNet & \texttt{param=shufflenet\_v2 shape=[224,224,3]} \\
 &  & ResNet     & \texttt{param=resnet18 shape=[224,224,3]} \\
\midrule

\multirow{3}{*}{\textbf{GROMACS}}
 & \multirow{3}{*}{\texttt{./gmx mdrun -nt 16}} & Nonbonded & \texttt{-s cutoff.tpr -nsteps 5000} \\
 &  & PME       & \texttt{-s pme.tpr -nsteps 5000} \\
 &  & LINCS     & \texttt{-s lincs.tpr -nsteps 5000} \\
\bottomrule
\end{tabular}%
}
\end{minipage}
\end{table*}

\section{Evaluation}
\subsection{Setup}

\textbf{Benchmarks.} We evaluate RepoOMP on workloads that vary substantially in structure and scale. We use NPB\cite{npb} and BOTS\cite{bots} as controlled benchmark suites to cover regular loop nests and irregular task-parallel patterns, respectively. These suites contain many comparatively small benchmark tasks and mainly serve as structured diagnostics for the rule branch, the agent branch, and their interaction under known kernels; the stronger repository-context evidence in this paper comes from FFmpeg\cite{ffmpeg}, NCNN\cite{ncnn}, and GROMACS, where cross-file dependency recovery, build integration, and large-code retrieval pressure are central. For these three large repositories, the paper reports three complementary views instead of a single exhaustive candidate matrix: hotspot-level funnel statistics over all reported hotspots, repository-level accepted-set distributions over all accepted real-world hotspots, and detailed matched-backbone performance tables for three accepted kernels per repository. The detailed kernels were selected to cover all three repositories and to include low-, middle-, and high-ranked accepted hotspots within each repository rather than only the fastest examples; Table~\ref{tab:accepted_case_rank} reports their within-repository positions. These ranks indicate that the detailed kernels are not exclusively the fastest accepted hotspots. In total, the accepted set contains 330 real-world repository hotspots across FFmpeg, NCNN, and GROMACS, and the paper reports both repository-level distribution summaries for that full accepted set and nine kernels expanded into detailed case studies. Table~\ref{tab:datasets} summarizes the resulting dataset statistics and the scale of the Multi-granularity Attributes Performance graph (MAP) constructed for each repository. The transition from micro-benchmarks to real-world applications substantially increases graph size, making repository-level dependency recovery increasingly central to the task. Table~\ref{tab:benchmarks_split} lists the kernels and execution configurations used in the evaluation.

The command fragments in Table~\ref{tab:benchmarks_split} also define the task interface used in the traditional-tool comparison: each real-world optimization begins from an executable command and a profiled repository rather than from a preselected source loop. We therefore interpret the traditional-tool results as end-to-end suitability results for this command-level hotspot-discovery protocol, not as an oracle-targeted comparison of transformation quality after a human has selected the exact loop or compilation unit. This interface is important because AutoPar/Polly-style tools operate once a target file or compilation unit is provided, whereas the repository-level task additionally requires discovering which files and loops should be optimized for the command and which edits preserve its observable checks. The command fragments serve as workload identifiers and reproducibility anchors. In the current artifact, the corresponding replay path consists of the repository command, the associated wrapper logic, the required model files or prepared inputs, and the runtime assets needed to execute the workload on repositories such as NCNN and GROMACS.

\begin{table*}[t]
  \centering
  \caption{Performance and cost evaluation on NPB executed with 16 threads on the Intel CPU. Speedup is relative to \texttt{-O3} serial execution, and token usage is reported in thousands. \textit{WA} indicates wrong answers, \textit{BF} indicates build failures, and \textit{NO} marks inapplicable token costs. \textbf{Avg.} represents the arithmetic mean across all benchmarks, treating WA/BF speedup as 0; we interpret the averages jointly with the per-task BF/WA breakdown rather than as standalone summaries. The aggregate rows show the average of their respective models. Each split table repeats Avg. for readability. In the method names, GPT abbreviates GPT-5.1, Claude abbreviates Claude 4.5 Sonnet, and Gemini abbreviates Gemini 3 Pro.}
  \label{tab:results_speedup_token}
  \EvalTableStyle{1.7pt}
  \begin{tabular*}{\textwidth}{L{0.21\textwidth} @{\extracolsep{\fill}} cc cccccccccc}
    \toprule
    \multirow{2}{*}{Method} & \multicolumn{2}{c}{\textbf{Avg.}} & \multicolumn{2}{c}{BT} & \multicolumn{2}{c}{CG} & \multicolumn{2}{c}{EP} & \multicolumn{2}{c}{FT} & \multicolumn{2}{c}{IS} \\
    \cmidrule(lr){2-3} \cmidrule(lr){4-5} \cmidrule(lr){6-7} \cmidrule(lr){8-9} \cmidrule(lr){10-11} \cmidrule(lr){12-13}
     & Spd. & Tok. & Spd. & Tok. & Spd. & Tok. & Spd. & Tok. & Spd. & Tok. & Spd. & Tok. \\
    \midrule
    ClaudeCode-based\cite{claudecode} & 3.31$\times$ & 1055.1 & \multicolumn{10}{c}{} \\
    \hspace{1em}+GPT\cite{gpt_2025} & 1.40$\times$ & 192.8 & \textit{WA} & 183.2 & 8.53$\times$ & 188.6 & \textit{WA} & 197.7 & \textit{BF} & 209.2 & 1.33$\times$ & 13.9 \\
    \hspace{1em}+Claude\cite{claude_2025} & 4.77$\times$ & 458.9 & 1.23$\times$ & 622.4 & 9.01$\times$ & 694.7 & 15.28$\times$ & 265.8 & \textit{WA} & 1520.2 & 3.27$\times$ & 9.2 \\
    \hspace{1em}+Gemini\cite{gemini_2025} & 3.77$\times$ & 2513.7 & \textit{WA} & 1349.1 & \textit{WA} & 1643.4 & 15.83$\times$ & 207.9 & 3.00$\times$ & 337.5 & 3.27$\times$ & 13.3 \\
    \addlinespace
    \textbf{RepoOMP (Ours)} & \textbf{8.23$\times$} & \textbf{76.7} & \multicolumn{10}{c}{} \\
    \hspace{1em}+GPT\cite{gpt_2025} & 8.12$\times$ & 76.6 & 10.23$\times$ & 92.4 & 10.32$\times$ & 18.2 & 15.78$\times$ & 30.1 & 10.73$\times$ & 34.9 & 1.37$\times$ & 20.5 \\
    \hspace{1em}+Claude\cite{claude_2025} & 7.31$\times$ & 46.6 & 9.73$\times$ & 69.7 & 9.87$\times$ & 11.6 & 13.33$\times$ & 24.0 & 10.64$\times$ & 25.5 & 1.39$\times$ & 7.7 \\
    \hspace{1em}+Gemini\cite{gemini_2025} & 9.26$\times$ & 106.8 & 11.30$\times$ & 134.6 & 12.31$\times$ & 24.8 & 15.95$\times$ & 46.6 & 11.57$\times$ & 67.0 & 3.33$\times$ & 28.3 \\
    \midrule
    AutoPar\cite{quinlan2011rose} & 1.53$\times$ & \textit{NO} & \textit{WA} & \textit{NO} & 0.05$\times$ & \textit{NO} & 7.83$\times$ & \textit{NO} & \textit{WA} & \textit{NO} & 3.08$\times$ & \textit{NO} \\
    Polly\cite{polly} & 1.37$\times$ & \textit{NO} & 1.12$\times$ & \textit{NO} & 1.05$\times$ & \textit{NO} & 1.77$\times$ & \textit{NO} & 0.58$\times$ & \textit{NO} & 1.00$\times$ & \textit{NO} \\
    Human Expert\cite{npb} & 6.25$\times$ & \textit{NO} & 9.48$\times$ & \textit{NO} & 5.57$\times$ & \textit{NO} & 14.50$\times$ & \textit{NO} & 2.93$\times$ & \textit{NO} & 3.27$\times$ & \textit{NO} \\
    \bottomrule
  \end{tabular*}
\end{table*}

\begin{table*}[t]
  \centering
  \caption{Performance and cost evaluation on NPB executed with 16 threads on the Intel CPU (continued). This continuation table reports the remaining benchmarks while repeating Avg. for readability. BF/WA speedup is treated as 0 in the average. In the method names, GPT abbreviates GPT-5.1, Claude abbreviates Claude 4.5 Sonnet, and Gemini abbreviates Gemini 3 Pro.}
  \label{tab:results_speedup_token_cont}
  \EvalTableStyle{1.9pt}
  \begin{tabular*}{\textwidth}{L{0.21\textwidth} @{\extracolsep{\fill}} cc cccccccccc}
    \toprule
    \multirow{2}{*}{Method} & \multicolumn{2}{c}{\textbf{Avg.}} & \multicolumn{2}{c}{LU} & \multicolumn{2}{c}{MG} & \multicolumn{2}{c}{SP} \\
    \cmidrule(lr){2-3} \cmidrule(lr){4-5} \cmidrule(lr){6-7} \cmidrule(lr){8-9}
     & Spd. & Tok. & Spd. & Tok. & Spd. & Tok. & Spd. & Tok. \\
    \midrule
    ClaudeCode-based\cite{claudecode} & 3.31$\times$ & 1055.1 & \multicolumn{6}{c}{} \\
    \hspace{1em}+GPT\cite{gpt_2025} & 1.40$\times$ & 192.8 & 0.74$\times$ & 256.3 & 0.60$\times$ & 229.4 & \textit{WA} & 263.7 \\
    \hspace{1em}+Claude\cite{claude_2025} & 4.77$\times$ & 458.9 & 0.90$\times$ & 103.2 & 7.61$\times$ & 113.9 & 0.87$\times$ & 341.7 \\
    \hspace{1em}+Gemini\cite{gemini_2025} & 3.77$\times$ & 2513.7 & \textit{BF} & 6042.2 & 6.54$\times$ & 984.1 & 1.55$\times$ & 9531.9 \\
    \addlinespace
    \textbf{RepoOMP (Ours)} & \textbf{8.23$\times$} & \textbf{76.7} & \multicolumn{6}{c}{} \\
    \hspace{1em}+GPT\cite{gpt_2025} & 8.12$\times$ & 76.6 & 1.13$\times$ & 134.5 & 11.73$\times$ & 193.2 & 3.67$\times$ & 89.0 \\
    \hspace{1em}+Claude\cite{claude_2025} & 7.31$\times$ & 46.6 & 1.17$\times$ & 101.1 & 9.09$\times$ & 89.4 & 3.25$\times$ & 43.6 \\
    \hspace{1em}+Gemini\cite{gemini_2025} & 9.26$\times$ & 106.8 & 3.51$\times$ & 170.9 & 11.86$\times$ & 240.6 & 4.25$\times$ & 141.9 \\
    \midrule
    AutoPar\cite{quinlan2011rose} & 1.53$\times$ & \textit{NO} & \textit{WA} & \textit{NO} & 0.19$\times$ & \textit{NO} & 1.05$\times$ & \textit{NO} \\
    Polly\cite{polly} & 1.37$\times$ & \textit{NO} & 3.81$\times$ & \textit{NO} & 0.90$\times$ & \textit{NO} & 0.72$\times$ & \textit{NO} \\
    Human Expert\cite{npb} & 6.25$\times$ & \textit{NO} & 8.80$\times$ & \textit{NO} & 4.49$\times$ & \textit{NO} & 0.98$\times$ & \textit{NO} \\
    \bottomrule
  \end{tabular*}
\end{table*}

\begin{table*}[t]
  \centering
  \caption{Performance and cost evaluation on BOTS executed with 16 threads on the Intel CPU. Speedup is relative to \texttt{-O3} serial execution, and token usage is reported in thousands. \textit{WA} indicates wrong answers, \textit{BF} indicates build failures, and \textit{NO} marks inapplicable token costs. \textbf{Avg.} represents the arithmetic mean across all benchmarks, treating WA/BF speedup as 0. The aggregate rows show the average of their respective models. Each split table repeats Avg. for readability. In the method names, GPT abbreviates GPT-5.1, Claude abbreviates Claude 4.5 Sonnet, and Gemini abbreviates Gemini 3 Pro.}
  \label{tab:bots_results_speedup_token}

  \EvalTableStyle{1.3pt}

  \begin{tabular*}{\textwidth}{L{0.21\textwidth} @{\extracolsep{\fill}} cc cccccccccc}
    \toprule
    \multirow{2}{*}{Method} & \multicolumn{2}{c}{\textbf{Avg.}} & \multicolumn{2}{c}{Alignment} & \multicolumn{2}{c}{FFT} & \multicolumn{2}{c}{Floorplan} & \multicolumn{2}{c}{Health} & \multicolumn{2}{c}{NQueens} \\
    \cmidrule(lr){2-3} \cmidrule(lr){4-5} \cmidrule(lr){6-7} \cmidrule(lr){8-9} \cmidrule(lr){10-11} \cmidrule(lr){12-13}
     & Spd. & Tok. & Spd. & Tok. & Spd. & Tok. & Spd. & Tok. & Spd. & Tok. & Spd. & Tok. \\
    \midrule

    ClaudeCode-based\cite{claudecode} & 4.12$\times$ & 611.3 & \multicolumn{10}{c}{} \\
    \hspace{1em}+GPT\cite{gpt_2025} & 3.01$\times$ & 244.2 & \textit{WA} & 344.1 & \textit{BF} & 742.5 & \textit{WA} & 203.2 & 0.21$\times$ & 163.5 & \textit{WA} & 125.2 \\
    \hspace{1em}+Claude\cite{claude_2025} & 1.82$\times$ & 430.7 & \textit{WA} & 480.6 & 2.20$\times$ & 866.0 & \textit{WA} & 268.9 & \textit{BF} & 506.2 & \textit{BF} & 259.2 \\
    \hspace{1em}+Gemini\cite{gemini_2025} & 7.52$\times$ & 1159.1 & \textit{BF} & 910.7 & 1.37$\times$ & 1310.2 & 1.95$\times$ & 910.5 & 10.10$\times$ & 361.3 & \textit{WA} & 705.9 \\
    \addlinespace

    \textbf{RepoOMP (Ours)} & \textbf{8.96$\times$} & \textbf{65.9} & \multicolumn{10}{c}{} \\
    \hspace{1em}+GPT\cite{gpt_2025} & 8.33$\times$ & 72.2 & 9.48$\times$ & 95.5 & 6.24$\times$ & 113.5 & 8.85$\times$ & 76.6 & 8.98$\times$ & 51.1 & 9.36$\times$ & 79.6 \\
    \hspace{1em}+Claude\cite{claude_2025} & 9.71$\times$ & 50.6 & 12.39$\times$ & 65.8 & 6.81$\times$ & 91.8 & 9.88$\times$ & 54.7 & 9.83$\times$ & 30.3 & 12.58$\times$ & 47.1 \\
    \hspace{1em}+Gemini\cite{gemini_2025} & 8.85$\times$ & 74.8 & 9.59$\times$ & 98.4 & 8.40$\times$ & 123.4 & 8.08$\times$ & 78.5 & 8.94$\times$ & 45.3 & 9.77$\times$ & 77.4 \\
    \midrule

    AutoPar\cite{quinlan2011rose} & 0.09$\times$ & \textit{NO} & \textit{WA} & \textit{NO} & \textit{WA} & \textit{NO} & 0.04$\times$ & \textit{NO} & \textit{WA} & \textit{NO} & \textit{WA} & \textit{NO} \\
    Polly\cite{polly} & 1.04$\times$ & \textit{NO} & 0.38$\times$ & \textit{NO} & 1.43$\times$ & \textit{NO} & 0.94$\times$ & \textit{NO} & 1.23$\times$ & \textit{NO} & 1.02$\times$ & \textit{NO} \\
    Human Expert\cite{bots} & 7.57$\times$ & \textit{NO} & 10.96$\times$ & \textit{NO} & 1.63$\times$ & \textit{NO} & 10.47$\times$ & \textit{NO} & 8.83$\times$ & \textit{NO} & 4.72$\times$ & \textit{NO} \\
    \bottomrule
  \end{tabular*}
\end{table*}

\begin{table*}[t]
  \centering
  \caption{Performance and cost evaluation on BOTS executed with 16 threads on the Intel CPU (continued). This continuation table reports the remaining benchmarks while repeating Avg. for readability. In the method names, GPT abbreviates GPT-5.1, Claude abbreviates Claude 4.5 Sonnet, and Gemini abbreviates Gemini 3 Pro.}
  \label{tab:bots_results_speedup_token_cont}

  \EvalTableStyle{1.5pt}

  \begin{tabular*}{\textwidth}{L{0.21\textwidth} @{\extracolsep{\fill}} cc cccccccccc}
    \toprule
    \multirow{2}{*}{Method} & \multicolumn{2}{c}{\textbf{Avg.}} & \multicolumn{2}{c}{Sort} & \multicolumn{2}{c}{SparseLU} & \multicolumn{2}{c}{Strassen} \\
    \cmidrule(lr){2-3} \cmidrule(lr){4-5} \cmidrule(lr){6-7} \cmidrule(lr){8-9}
     & Spd. & Tok. & Spd. & Tok. & Spd. & Tok. & Spd. & Tok. \\
    \midrule

    ClaudeCode-based\cite{claudecode} & 4.12$\times$ & 611.3 & \multicolumn{6}{c}{} \\
    \hspace{1em}+GPT\cite{gpt_2025} & 3.01$\times$ & 244.2 & 6.79$\times$ & 120.1 & 11.67$\times$ & 121.0 & 5.37$\times$ & 134.3 \\
    \hspace{1em}+Claude\cite{claude_2025} & 1.82$\times$ & 430.7 & 0.96$\times$ & 390.9 & 11.19$\times$ & 285.0 & 0.22$\times$ & 389.1 \\
    \hspace{1em}+Gemini\cite{gemini_2025} & 7.52$\times$ & 1159.1 & 1.02$\times$ & 1526.7 & 38.57$\times$ & 313.1 & 7.17$\times$ & 3234.4 \\
    \addlinespace

    \textbf{RepoOMP (Ours)} & \textbf{8.96$\times$} & \textbf{65.9} & \multicolumn{6}{c}{} \\
    \hspace{1em}+GPT\cite{gpt_2025} & 8.33$\times$ & 72.2 & 9.54$\times$ & 66.9 & 8.02$\times$ & 54.3 & 6.15$\times$ & 39.9 \\
    \hspace{1em}+Claude\cite{claude_2025} & 9.71$\times$ & 50.6 & 9.14$\times$ & 58.7 & 11.13$\times$ & 35.7 & 5.93$\times$ & 20.4 \\
    \hspace{1em}+Gemini\cite{gemini_2025} & 8.85$\times$ & 74.8 & 7.81$\times$ & 84.0 & 10.89$\times$ & 46.0 & 7.30$\times$ & 45.1 \\
    \midrule

    AutoPar\cite{quinlan2011rose} & 0.09$\times$ & \textit{NO} & 0.66$\times$ & \textit{NO} & \textit{WA} & \textit{NO} & \textit{WA} & \textit{NO} \\
    Polly\cite{polly} & 1.04$\times$ & \textit{NO} & 1.00$\times$ & \textit{NO} & 1.17$\times$ & \textit{NO} & 1.13$\times$ & \textit{NO} \\
    Human Expert\cite{bots} & 7.57$\times$ & \textit{NO} & 6.63$\times$ & \textit{NO} & 11.61$\times$ & \textit{NO} & 5.73$\times$ & \textit{NO} \\
    \bottomrule
  \end{tabular*}
\end{table*}

\begin{table*}[t]
  \centering
  \caption{Performance and token cost on the real-world applications (FFmpeg and NCNN) at 16 threads on Intel. \textbf{Avg.} denotes the arithmetic mean across all nine sub-tasks. Each real-world cell is the arithmetic mean over five repeated executions of the same generated implementation under the same workload and reported 16-thread configuration. Speedup is relative to serial \texttt{-O3}; token usage is reported in thousands (k).}
  \label{tab:large_repo_results}

  \tiny
  \captionsetup{skip=6pt}
  \renewcommand{\arraystretch}{1.12}
  \setlength{\tabcolsep}{0.2pt}

  \begin{tabular*}{\textwidth}{L{0.165\textwidth} @{\extracolsep{\fill}} cc cccccccccccc}
    \toprule
    \multirow{2}{*}{Method} & \multicolumn{2}{c}{\textbf{Avg.}} & \multicolumn{6}{c}{FFmpeg} & \multicolumn{6}{c}{NCNN} \\
    \cmidrule(lr){2-3} \cmidrule(lr){4-9} \cmidrule(lr){10-15}
     & Spd. & Tok. & \multicolumn{2}{c}{MPEG4} & \multicolumn{2}{c}{NLMeans} & \multicolumn{2}{c}{Deshake} & \multicolumn{2}{c}{MobNet} & \multicolumn{2}{c}{ShuffNet} & \multicolumn{2}{c}{ResNet} \\
    \cmidrule(lr){4-5} \cmidrule(lr){6-7} \cmidrule(lr){8-9} \cmidrule(lr){10-11} \cmidrule(lr){12-13} \cmidrule(lr){14-15}
     & & & Spd. & Tok. & Spd. & Tok. & Spd. & Tok. & Spd. & Tok. & Spd. & Tok. & Spd. & Tok. \\
    \midrule

    ClaudeCode/Cl.\cite{claude_2025} & 4.17$\times$ & 111.1 & 0.80$\times$ & 95.2 & 1.01$\times$ & 80.4 & 1.39$\times$ & 100.8 & 5.35$\times$ & 55.5 & 4.15$\times$ & 337.3 & 8.72$\times$ & 112.7 \\
    \addlinespace

    RepoOMP/GPT\cite{gpt_2025} & 5.34$\times$ & 58.7 & 1.13$\times$ & 65.9 & 5.29$\times$ & 59.8 & 1.94$\times$ & 50.3 & 5.65$\times$ & 7.3 & 5.02$\times$ & 208.9 & 9.91$\times$ & 49.4 \\
    RepoOMP/Cl.\cite{claude_2025} & 4.94$\times$ & 50.6 & 1.05$\times$ & 58.4 & 1.17$\times$ & 42.6 & 2.02$\times$ & 39.5 & 6.05$\times$ & 6.6 & 4.12$\times$ & 189.4 & 9.99$\times$ & 48.9 \\
    RepoOMP/Gem.\cite{gemini_2025} & 5.46$\times$ & 66.4 & 1.15$\times$ & 78.2 & 5.53$\times$ & 69.9 & 2.04$\times$ & 62.0 & 5.93$\times$ & 7.8 & 4.47$\times$ & 223.6 & 9.37$\times$ & 52.8 \\
    \bottomrule
  \end{tabular*}
\end{table*}

\begin{table*}[t]
  \centering
  \caption{Performance and token cost on the real-world applications (continued with GROMACS). Each real-world cell is the arithmetic mean over five repeated executions of the same generated implementation under the same workload and reported 16-thread configuration. Speedup is relative to serial \texttt{-O3}; token usage is reported in thousands (k).}
  \label{tab:large_repo_results_cont}

  \EvalTableStyle{2.0pt}

  \begin{tabular*}{\textwidth}{L{0.19\textwidth} @{\extracolsep{\fill}} cc cccccccccc}
    \toprule
    \multirow{2}{*}{Method} & \multicolumn{2}{c}{\textbf{Avg.}} & \multicolumn{6}{c}{GROMACS} \\
    \cmidrule(lr){2-3} \cmidrule(lr){4-9}
     & Spd. & Tok.(k) & \multicolumn{2}{c}{Nonbd} & \multicolumn{2}{c}{PME} & \multicolumn{2}{c}{LINCS} \\
    \cmidrule(lr){4-5} \cmidrule(lr){6-7} \cmidrule(lr){8-9}
     & & & Spd. & Tok. & Spd. & Tok. & Spd. & Tok. \\
    \midrule

    ClaudeCode/Cl.\cite{claude_2025} & 4.17$\times$ & 111.1 & 7.25$\times$ & 109.3 & 3.47$\times$ & 53.3 & 5.43$\times$ & 55.5 \\
    \addlinespace

    RepoOMP/GPT\cite{gpt_2025} & 5.34$\times$ & 58.7 & 8.23$\times$ & 23.9 & 4.52$\times$ & 28.2 & 6.40$\times$ & 35.0 \\
    RepoOMP/Cl.\cite{claude_2025} & 4.94$\times$ & 50.6 & 7.93$\times$ & 18.4 & 5.19$\times$ & 20.8 & 6.98$\times$ & 30.8 \\
    RepoOMP/Gem.\cite{gemini_2025} & 5.46$\times$ & 66.4 & 8.30$\times$ & 32.6 & 4.68$\times$ & 30.9 & 7.71$\times$ & 39.8 \\
    \bottomrule
  \end{tabular*}
\end{table*}

\begin{table*}[t]
  \centering
  \caption{Repository-level accepted-set outer-frame summary for the reported real-world repositories. Counts, acceptance rates, and coarse speedup bins are computed over the full accepted real-world sets, not a small audit slice. Att. denotes attempted transformations, Acc. denotes accepted hotspots, Acc.(P/A) denotes acceptance over profiled and attempted hotspots, Med. denotes median speedup, and IQR denotes the interquartile speedup range.}
  \label{tab:realworld_summary}
  \scriptsize
  \renewcommand{\arraystretch}{1.12}
  \setlength{\tabcolsep}{2.8pt}
  \resizebox{\textwidth}{!}{%
    \begin{tabular}{ccccccccccc}
      \toprule
      Repo & Profiled & Att. & Acc. & Acc.(P/A) & Below 1.2$\times$ & 1.2--1.5$\times$ & 1.5--2.0$\times$ & At least 2.0$\times$ & Med. & IQR \\
      \midrule
      FFmpeg & 552 & 461 & 221 & 40.0/47.9\% & 30 & 60 & 50 & 81 & 1.85x & 1.35--3.20x \\
      NCNN & 147 & 114 & 40 & 27.2/35.1\% & 2 & 5 & 8 & 25 & 4.50x & 2.10--6.80x \\
      GROMACS & 163 & 137 & 69 & 42.3/50.4\% & 5 & 10 & 15 & 39 & 3.60x & 1.80--6.20x \\
      \midrule
      Total RW & 862 & 712 & 330 & 38.3/46.3\% & 37 & 75 & 73 & 145 & 2.25x & 1.45--4.80x \\
      \bottomrule
    \end{tabular}
  }
\end{table*}

\begin{table}[t]
  \centering
  \caption{Per-kernel 16-thread repeated-run speedup disclosure for the nine reported real-world case studies. Each entry reports mean speedup with min--max range and coefficient of variation (CV) over five repeated executions of the same generated implementation under the same workload and Intel/\texttt{-O3}/16-thread setting. Claude Code denotes the same-kernel unstructured repository-agent result produced with Claude 4.5 Sonnet under the protocol in Tables~\ref{tab:large_repo_results}--\ref{tab:large_repo_results_cont}.}
  \label{tab:realworld_dispersion_detail}
  \footnotesize
  \renewcommand{\arraystretch}{1.06}
  \setlength{\tabcolsep}{4pt}
  \begin{tabular}{ccc}
    \toprule
    Kernel & RepoOMP & Claude Code \\
    \midrule
    FF-MPEG4 & 1.13x [1.12, 1.14], 0.64\% & 0.80x [0.76, 0.84], 5.23\% \\
    FF-NLMeans & 5.29x [5.26, 5.32], 0.57\% & 1.01x [0.93, 1.08], 7.15\% \\
    FF-Deshake & 2.02x [2.02, 2.03], 0.23\% & 1.39x [1.32, 1.46], 4.82\% \\
    NCNN-MobNet & 6.05x [6.01, 6.10], 0.56\% & 5.35x [5.00, 5.70], 6.45\% \\
    NCNN-ShuffNet & 4.47x [4.44, 4.50], 0.50\% & 4.15x [3.90, 4.40], 5.91\% \\
    NCNN-ResNet & 9.91x [9.86, 9.95], 0.37\% & 8.72x [8.03, 9.41], 7.88\% \\
    GMX-Nonbd & 8.30x [8.24, 8.35], 0.49\% & 7.25x [6.93, 7.56], 4.34\% \\
    GMX-PME & 5.19x [5.13, 5.24], 0.92\% & 3.47x [3.23, 3.70], 6.72\% \\
    GMX-LINCS & 6.40x [6.33, 6.45], 0.74\% & 5.43x [5.12, 5.73], 5.56\% \\
    \bottomrule
  \end{tabular}
\end{table}

%
%

\begin{table}[t]
  \centering
  \caption{Hotspot-level funnel and end-to-end cost on the reported workloads. \textit{Hotspots} denotes profiled candidate hotspots. \textit{Route(H/M/L)} gives the high-, medium-, and low-confidence routing counts before transformation. \textit{Attempted} counts candidates for which RepoOMP attempted a source transformation. \textit{Outcome(CF/Chk/RB/Acc)} denotes compilation failures, workload-check failures, no-speedup rollbacks, and accepted hotspots. \textit{Acc.(P/A)\%} reports acceptance over profiled hotspots and over attempted transformations. \textit{Total(s)} and \textit{Token(k)} are average end-to-end wall-clock cost and model-side token usage.}
  \label{tab:candidate_selection}
  \small
  \renewcommand{\arraystretch}{1.08}
  \setlength{\tabcolsep}{2.2pt}
  \resizebox{\columnwidth}{!}{%
  \begin{tabular}{c c c c c c c c}
    \toprule
    Repo & Hotspots & Route(H/M/L) & Attempted & Outcome(CF/Chk/RB/Acc) & Acc.(P/A)\% & Total(s) & Token(k) \\
    \midrule
    NPB      & 78  & 5/62/11    & 67  & 7/9/17/34     & 43.6/50.7 & 23.1  & 76.7 \\
    BOTS     & 11  & 1/9/1      & 10  & 1/1/0/8       & 72.7/80.0 & 28.6  & 65.9 \\
    FFmpeg   & 552 & 33/428/91  & 461 & 80/60/100/221 & 40.0/47.9 & 139.2 & 58.7 \\
    NCNN     & 147 & 11/103/33  & 114 & 15/25/34/40   & 27.2/35.1 & 97.5  & 88.5 \\
    GROMACS  & 163 & 10/127/26  & 137 & 20/15/36/69   & 42.3/50.4 & 118.4 & 29.0 \\
    \bottomrule
  \end{tabular}
  }
\end{table}

\begin{table}[t]
  \centering
  \caption{ThreadSanitizer audit on the nine reported real-world workloads. \textit{Orig.} and \textit{RepoOMP} denote observed warning counts under the same Intel/\texttt{-O3}/16-thread command template, \textit{Delta\%} is the relative change from Orig. to RepoOMP, and \textit{Owner} is a short attribution-boundary label from manual inspection.}
  \label{tab:tsan_summary}
  \footnotesize
  \renewcommand{\arraystretch}{1.08}
  \setlength{\tabcolsep}{4pt}
  \begin{tabular}{cccccc}
    \toprule
    Repo & Workload & Orig. & RepoOMP & Delta\% & Owner \\
    \midrule
    FFmpeg & MPEG4      & 39    & 20   & -48.7\% & Mixed \\
    FFmpeg & NLMeans    & 17    & 11   & -35.3\% & Mixed \\
    FFmpeg & Deshake    & 18    & 13   & -27.8\% & Mixed \\
    NCNN   & MobileNet  & 1627  & 1260 & -22.6\% & Runtime \\
    NCNN   & ShuffleNet & 10453 & 8195 & -21.6\% & Mixed \\
    NCNN   & ResNet     & 1178  & 639  & -45.8\% & Mixed \\
    GROMACS & Nonbd     & 639   & 362  & -43.3\% & Mixed \\
    GROMACS & PME       & 793   & 473  & -40.4\% & Mixed \\
    GROMACS & LINCS     & 858   & 520  & -39.4\% & Mixed \\
    \bottomrule
  \end{tabular}
\end{table}

\textbf{Comparison Methods.} We compare RepoOMP against three baselines that represent distinct automation paradigms. For traditional automatic parallelization, we use AutoPar as a source-to-source baseline and Polly to represent polyhedral optimization within LLVM. This selection covers the main design space of compiler-driven OpenMP insertion while excluding systems that require substantial manual intervention, such as dynamic frameworks with interactive guidance. For the generative baseline, we use Claude Code to represent repository-scale agentic optimization. In the evaluated artifact, the Claude Code baseline uses the npm package v1.0.8. In our implementation of this baseline, the agent operates without MAP-guided routing, hotspot summaries, or Structured Transformation Context (STC) compression; it therefore acts as a matched unstructured repository-agent baseline. More general repository agents such as SWE-agent, OpenHands, and Devin are discussed in Section~2.3 as adjacent systems; the real-world comparison here focuses on hotspot-local OpenMP transformation under the matched retry-budget protocol used throughout this paper. On the real-world repositories, our primary comparison is same-hotspot and same-retry-budget against this AI-agent baseline. Traditional automatic parallelizers remain in the benchmark-suite comparison, while the real-world tables center on the agent baseline because FFmpeg, NCNN, and GROMACS require repository-specific build macros, cross-translation-unit headers, and workload wrappers that make per-hotspot compiler-tool integration materially different from the matched agent protocol.

We interpret baselines through task applicability rather than a single interchangeable leaderboard: AutoPar/Polly are structured-suite local baselines requiring a file, loop, or compilation unit; Rule-Only exposes local-rule coverage after MAP routing; Claude Code is the matched real-world agent baseline without MAP/STC; and RepoOMP targets the full command-level input of workload command, profile, and repository.

\textbf{Backbone Models.} We use three general-purpose foundation models as the optimization backbones in RepoOMP: Claude 4.5 Sonnet\cite{claude_2025}, Gemini 3 Pro\cite{gemini_2025}, and GPT-5.1\cite{gpt_2025}. Repository-scale OpenMP optimization depends on cross-file reasoning over transitive dependencies, shared state, and validation feedback, so the main system is built around these general-purpose backbones paired with MAP. Specialized HPC pretraining and graph-based inference remain natural extensions of the same workflow.

\textbf{Baseline Protocol.} To ensure fair comparison, the agent baselines operate under a matched protocol. Each baseline agent receives the same hotspot location and repository access as RepoOMP, but without MAP summaries, router decisions, or STC compression. Both RepoOMP and the baselines operate under an identical retry budget of three attempts per hotspot: after three failed compilation or validation cycles, the task is recorded as BF or WA and no further generation is attempted for that candidate. Token accounting captures cumulative model-side usage across all attempts within the retry budget, including failed generations, so that the cost of trial-and-error behavior is reflected in the reported totals. Repository-indexing overhead (e.g., RAG retrieval, file scanning) is excluded because it is common to all agents and not billable to the LLM endpoint.

For each reported task-level comparison, all compared methods are evaluated under the same workload input, same 16-thread target setting, and the same -O3 serial reference used to compute speedup. The resulting pairwise differences are therefore configuration-matched comparisons within the same task.

On the real-world tables, we prioritize \textbf{matched-workload pairwise comparisons} as the primary evidence, with RepoOMP variants compared against the Claude Code baseline on the same accepted real-world kernels and retry budget. This layout keeps the comparison protocol tied to the configurations reported in the table while avoiding unsupported cross-task aggregation.

\textbf{Platforms.} We evaluate on two hardware platforms to assess whether the observed trends remain stable across architectures. The first platform uses dual-socket Intel Xeon Gold 6430 processors with 64 physical cores and 128 hardware threads at up to 3.40 GHz. The second uses a single-socket AMD EPYC 9654 processor with 96 cores and 192 hardware threads at 2.40 GHz. All artifacts are compiled on Linux with Clang 22.0.0, using \texttt{-O3} and OpenMP 4.5 support unless stated otherwise.

\textbf{Scope and Aggregation.} The paper reports four non-interchangeable evidence units in the main body. Tables~\ref{tab:results_speedup_token}, \ref{tab:bots_results_speedup_token}, \ref{tab:large_repo_results}, and \ref{tab:large_repo_results_cont} are \emph{task-level} results; for the real-world kernels, each cell is a five-run arithmetic mean under the same workload and reported 16-thread configuration. Table~\ref{tab:candidate_selection} is the \emph{hotspot-level} workflow audit, reporting routed counts, attempted transformations, final outcomes, and end-to-end wall-clock cost. Table~\ref{tab:realworld_summary} provides the \emph{repository-level} outer frame over all 330 accepted real-world hotspots by pairing acceptance counts/rates with full-set speedup bins and interquartile summaries, and Table~\ref{tab:realworld_dispersion_detail} provides the \emph{kernel-level} repeated-run disclosure for the nine case studies. Table~\ref{tab:tsan_summary} is a supplementary workload-level dynamic audit. Tables~\ref{tab:accepted_distribution_full} and \ref{tab:accepted_case_rank} extend the repository-level audit with fuller percentile disclosure and within-repository rank positions for the nine detailed kernels.

The real-world evaluation follows a hotspot-oriented funnel. Across the five workloads, 951 profiled hotspots exceed the 5\% runtime threshold, 817 enter attempted transformation, and 372 are accepted after compilation, workload-specific checks, and positive speedup. For the 712 attempted real-world transformations, Table~\ref{tab:ablation_funnel} reports 115 compilation failures, 110 workload-check failures, 157 no-speedup rollbacks, and 330 acceptances. Of these accepted hotspots, 330 belong to FFmpeg, NCNN, and GROMACS, while 42 belong to NPB/BOTS. The nine detailed real-world kernels are matched-backbone case studies drawn from that larger accepted real-world set. They were chosen to span repository diversity and the accepted-speedup range visible in the disclosed distributions, but they are not used as a random sample of all accepted transformations. Table~\ref{tab:realworld_summary} reports repository-level acceptance counts, rates, and speedup bins over the full accepted real-world sets; Tables~\ref{tab:accepted_distribution_full} and \ref{tab:accepted_case_rank} add percentile summaries and within-repository positions; Table~\ref{tab:realworld_dispersion_detail} discloses per-kernel five-run spread; and Table~\ref{tab:candidate_selection} complements token cost with end-to-end wall-clock accounting. Together, these views connect repository-level distributions to detailed matched-baseline kernel studies.

\subsection{RQ1: How effective is RepoOMP under the reported checks?}
Table~\ref{tab:candidate_selection} provides the hotspot-level workflow audit, while Tables~\ref{tab:results_speedup_token}, \ref{tab:bots_results_speedup_token}, \ref{tab:large_repo_results}, and \ref{tab:large_repo_results_cont} provide task-level case studies. Together, these views characterize RepoOMP's practical effectiveness under the reported protocol. RepoOMP reaches average speedups of $8.23\times$ on NPB, $8.96\times$ on BOTS, and $5.25\times$ on the nine reported accepted real-world kernels. On the reported real-world comparisons, RepoOMP variants improve the average from the Claude Code baseline's $4.17\times$ to $4.94$--$5.46\times$.

Among accepted real-world hotspots, the repository-level evidence shows that these nine kernels are not isolated wins. Tables~\ref{tab:realworld_summary}, \ref{tab:accepted_distribution_full}, and \ref{tab:accepted_case_rank} make the denominator explicit, disclose the accepted-set distribution, and place the detailed kernels inside it. FFmpeg has the broadest low-to-mid spread ($P25=1.35\times$, median $1.85\times$), while NCNN and GROMACS are more right-shifted (medians $4.50\times$ and $3.60\times$) but still retain low-end accepted cases. Across all 330 accepted real-world hotspots, the median is $2.25\times$, the interquartile range is $1.45$--$4.80\times$, and 145 hotspots are at least $2.0\times$. The same distribution remains substantial under stricter acceptance thresholds: 293/330 accepted hotspots stay at or above $1.2\times$, and 218/330 stay at or above $1.5\times$. The rank table further shows that the nine detailed kernels span repository-internal positions from FFmpeg-MPEG4 at $195/221$ to NCNN-ResNet at $3/40$, not only the top-ranked wins.

The stability evidence points in the same direction. Table~\ref{tab:realworld_dispersion_detail} reports the matched Claude 4.5 Sonnet source data for Claude Code and shows that the matched-backbone winner remains unchanged on all nine five-run comparisons. RepoOMP's 16-thread dispersion stays below 1\% CV on every reported kernel, while the Claude Code runs in this audit range from 4.34\% to 7.88\% CV. Under the executable checks used in this study, RepoOMP records zero BF/WA cells in the reported task tables, whereas the ClaudeCode-based workflows incur 26 such cells. The reported repository optimizations are useful only when a transformation survives compilation, validation, and execution under the tested configurations.

Table~\ref{tab:tsan_summary} adds a ThreadSanitizer comparison on the same nine real-world workloads. Observed warning counts are lower for RepoOMP in all nine matched audits. Using one matched 16-thread command template keeps the dynamic-race audit comparable across methods, while the `Owner' column localizes each warning to its repository context rather than leaving the analysis at the level of aggregate counts.

The choice of LLM affects the attainable performance ceiling, and the observed results suggest that MAP-guided evidence reduces, rather than eliminates, model sensitivity in the reported setting. Within RepoOMP, Gemini 3 Pro gives the highest average speedup on NPB ($9.26\times$) and the real-world tasks ($5.46\times$), Claude 4.5 is strongest on BOTS ($9.71\times$), and GPT-5.1 remains close on the real-world workloads ($5.34\times$). All three RepoOMP variants show zero BF/WA cells in the reported tasks, and even the lowest-cost configuration, RepoOMP+Claude 4.5, reaches $4.94\times$ on the real-world workloads.

The token results support the same interpretation. On the reported real-world hotspot workloads, RepoOMP reduces agent-side token usage by 47--68\% relative to the ClaudeCode-based baseline in the reported matched-workload comparisons. RepoOMP achieves these gains while keeping agent-side context compact. The results are consistent with the core argument of the paper: once dependency evidence is recovered, routed, and compressed before generation, both agent-side efficiency and validation stability improve in this evaluation.

\subsection{RQ2: What explains RepoOMP's gains over the selected baselines on the reported tasks?}
On the reported tasks, the results are consistent with the interpretation that RepoOMP benefits from addressing two evidence-access mismatches identified in Section~1. Rule-only tools mainly lose opportunities through conservative under-parallelization: they are safe when the relevant evidence is local and analyzable, but they often reject profitable loops when aliasing, cross-function side effects, or irregular control flow prevent a local proof. Agent-only tools fail in the opposite direction: they can attempt bolder transformations, but once the prompt misses a transitive dependency or contains too much irrelevant code, they become unstable, expensive to repair, and sometimes unsafe.

The tables make this asymmetry visible. AutoPar and Polly remain conservative and achieve relatively low speedups in this protocol, such as $1.53\times$ and $1.37\times$ on NPB and $0.09\times$ and $1.04\times$ on BOTS. ClaudeCode-based workflows sometimes find strong improvements on individual kernels, but they also account for many BF and WA outcomes while consuming substantially more tokens. RepoOMP sits between these extremes: MAP recovers repository-level dependency evidence before generation, and the router then decides whether a candidate should go to rules, to the agent, or into a conservative low-confidence path. The reported behavior is consistent with RepoOMP avoiding many local-provability losses without inheriting the full cost of unstructured agent retrieval.

Table~\ref{tab:candidate_selection} further supports this explanation at the hotspot level. Across the 951 reported hotspots, only 60 are categorized as high confidence, whereas 729 are categorized as middle confidence. This imbalance is important: most repository hotspots are not simple enough to be discharged by rules alone, but they are still structured enough to remain within the bounded STC regime instead of being discarded upfront. The low-confidence branch covers 162 hotspots, and the accepted hotspot counts vary across workloads, from 34 on NPB and 8 on BOTS to 221 on FFmpeg, 40 on NCNN, and 69 on GROMACS. Table~\ref{tab:candidate_selection} also shows that the unaccepted routed cases split meaningfully across failure modes instead of collapsing into a single bucket: 123 compile failures, 110 workload-check failures, and 212 no-speedup rollbacks, together accounting for all 445 unaccepted attempted transformations. These counts are consistent with the intended division of labor in RepoOMP: rules handle a small set of structurally clear cases, the agent handles the dominant bounded-but-nontrivial cases, and the router explicitly filters out the remainder when the propagated blockers remain too strong.

The running example from Section~1 illustrates the same mechanism. A loop that calls a helper which eventually updates file-scope shared statistics is easy to mishandle. A rule-only system often refuses to transform it because the write is not provably harmless from local context. A pure agent given only the loop body may parallelize it and introduce a race. RepoOMP instead records the transitive write in MAP, propagates that risk to the caller, and invokes the agent only when the resulting candidate still has a manageable dependency boundary. When that happens, the agent receives the target loop together with the relevant helper summaries and shared-state facts, turning the task into bounded restructuring instead of blind speculation. RQ2 therefore supports the paper's design argument: the challenge is non-local evidence, the prior flaws are method-specific evidence mismatches, and the practical gain of RepoOMP is consistent with correcting those mismatches explicitly.

\subsection{RQ3: How does context reduction affect correctness and cost?}
RQ3 examines the second half of our main argument: if agent-based parallelization fails because generic retrieval does not recover parallelization-relevant evidence, then the right intervention is not simply to enlarge prompts but to construct a Structured Transformation Context (STC). The observed outcomes are consistent with MAP reducing both omission risk and distraction cost. Without structured reduction, an agent must choose between a narrow local prompt that risks hiding a crucial dependency and a broad repository prompt that increases noise, token cost, and verification churn.

\textbf{Token compression and validation stability.}
The token results quantify this effect directly. On the real-world applications, the Claude Code baseline averages 111.1k tokens, whereas RepoOMP averages 58.6k. Similar reductions appear on NPB and BOTS. Importantly, this prompt reduction is accompanied by strong task-level validation outcomes under the executable checks used here. RepoOMP is also the only compared family without BF or WA in the reported tables. Within this evaluation, context reduction is associated with fewer observed BF/WA outcomes while preserving the dependency evidence that matters for transformation validity and removing unrelated repository text.

The same conclusion holds by backbone. On the real-world repositories, MAP keeps RepoOMP within 50.6k--66.4k tokens across Claude 4.5, GPT-5.1, and Gemini 3 Pro, while Claude Code uses 111.1k. Under this compressed context budget, RepoOMP's average speedup varies only from $4.94\times$ to $5.46\times$, and none of the three backbones produces BF or WA on the reported tasks. Similar compression appears on NPB and BOTS.

\textbf{Failure-mode control.}
The extreme outliers are also informative. Without MAP, a code agent pays more than a linear penalty for larger prompts: after compilation or validation begins to fail, it can append new hypotheses, revisit loosely related files, and rewrite earlier decisions without recovering the actual dependency cause. On NPB, ClaudeCode+Gemini 3 Pro consumes 6042.2k tokens on LU and still ends in BF, spends 9531.9k tokens on SP for only $1.55\times$ speedup, and incurs WA on BT and CG while averaging 2513.7k tokens. These anomalies are consistent with deferred verification loops; MAP constrains the search space before generation so failed attempts do not expand into uncontrolled context drift.

The ablation study supports the same interpretation. Figures~\ref{fig:ablation_speedup} and \ref{fig:ablation_token_cost} show that removing STC or flattening MAP into a function-level view lowers accepted-hotspot rate and accepted-set speedup while increasing token cost and wall-clock time. These results indicate that repository-scale optimization needs structured evidence compression rather than simply larger prompts.

\subsection{RQ4: How does RepoOMP behave across scales and hardware settings?}
Within the reported sensitivity studies, RepoOMP remains effective as the optimization setting shifts from benchmark kernels to large repositories and from one hardware configuration to another. On NPB and BOTS, the main challenge is recognizing profitable parallel patterns under local complexity. On FFmpeg, NCNN, and GROMACS, the dominant challenge becomes isolating the right dependency boundary inside a much larger body of unrelated code. RepoOMP remains competitive across these reported settings and selected baselines; the ablations suggest that MAP and routing help by scaling with repository structure rather than token-window size.

The sensitivity studies provide a more detailed picture. Figures~\ref{fig:npb_speedup}, \ref{fig:bots_speedup}, and \ref{fig:real_world_speedup} show that performance generally improves as thread count grows, although some kernels saturate because of memory bandwidth or synchronization limits. Figure~\ref{fig:Compiler} shows that source-level parallelization and compiler optimization are complementary. RepoOMP remains beneficial under lower optimization levels, but the gain is larger on complex real-world code when backend optimizations can exploit the exposed parallel structure. Figures~\ref{fig:cpu_cc} and \ref{fig:cpu_RepoOMP} further show lower variation for RepoOMP than for the Claude Code baseline in the reported two-platform study, which is consistent with the interpretation that explicit dependency and hotspot evidence reduces reliance on loosely selected repository context.

\subsection{Ablation Study}
We analyze the behavior of RepoOMP along four dimensions. First, we compare the full system against five simplified variants to probe how the integrated workflow behaves when one design element is removed: \emph{No-Router}, which disables the original routing policy and sends candidates directly through the agent path; \emph{No-STC}, which removes the bounded Structured Transformation Context while retaining the rest of the pipeline; \emph{Flat-Ctx}, which weakens MAP to a flatter function-level representation; \emph{Rule-Only}, which disables the LLM branch; and \emph{Agent-Only}, which disables the rule branch. These comparisons characterize component roles within a fixed Intel/Gemini setting. Second, we vary thread counts across 16, 32, and 64 to study scalability (Figures~\ref{fig:npb_speedup}, \ref{fig:bots_speedup}, \ref{fig:real_world_speedup}). Third, we compare \texttt{-O0} and \texttt{-O3} to examine how source-level parallelization interacts with compiler optimization. Finally, we repeat the evaluation on Intel and AMD CPUs to assess cross-platform stability.

\subsubsection{\textbf{Component Analysis}}

\begin{figure}[t]
  \centering
  \begin{minipage}[t]{0.48\linewidth}
    \centering
    \includegraphics[width=0.94\linewidth]{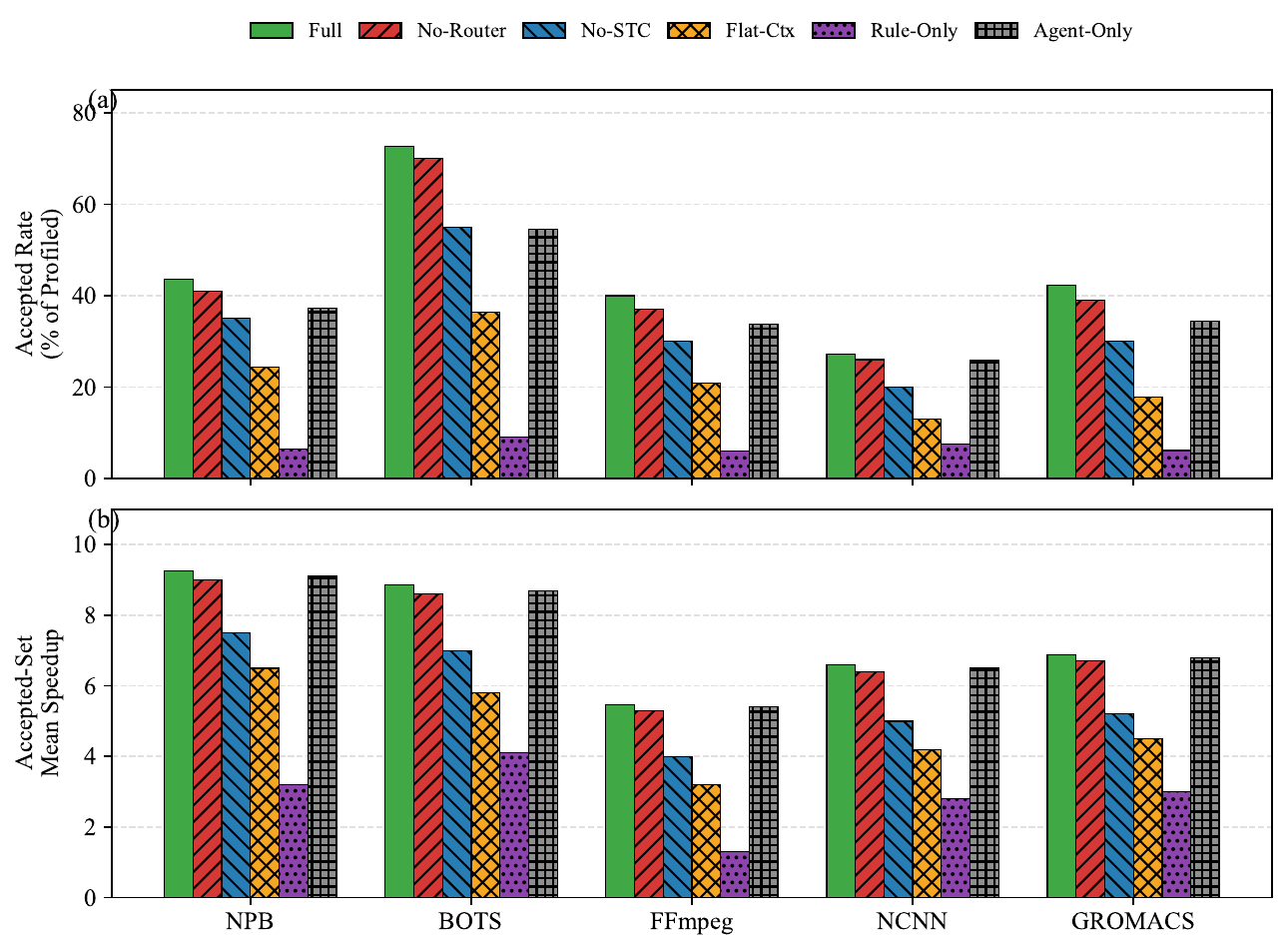}
    \vspace{0pt}
    \caption{Ablation study on RepoOMP's component effectiveness. Panel (a) reports accepted-hotspot rate, and panel (b) reports accepted-set mean speedup, for the full system and five ablated variants across the five workloads. All experiments were conducted on Intel with 16 threads, using \texttt{-O3} and Gemini 3.0 Pro.}
    \Description{A two-panel bar chart comparing the full RepoOMP pipeline against No-Router, No-STC, Flat-Ctx, Rule-Only, and Agent-Only across NPB, BOTS, FFmpeg, NCNN, and GROMACS. The top panel reports accepted-hotspot rate, and the bottom panel reports accepted-set mean speedup. The full system is strongest overall, Flat-Ctx and Rule-Only are notably weaker, and Agent-Only remains closer to the full system than Rule-Only.}
    \label{fig:ablation_speedup}
  \end{minipage}\hfill
  \begin{minipage}[t]{0.48\linewidth}
    \centering
    \includegraphics[width=0.94\linewidth]{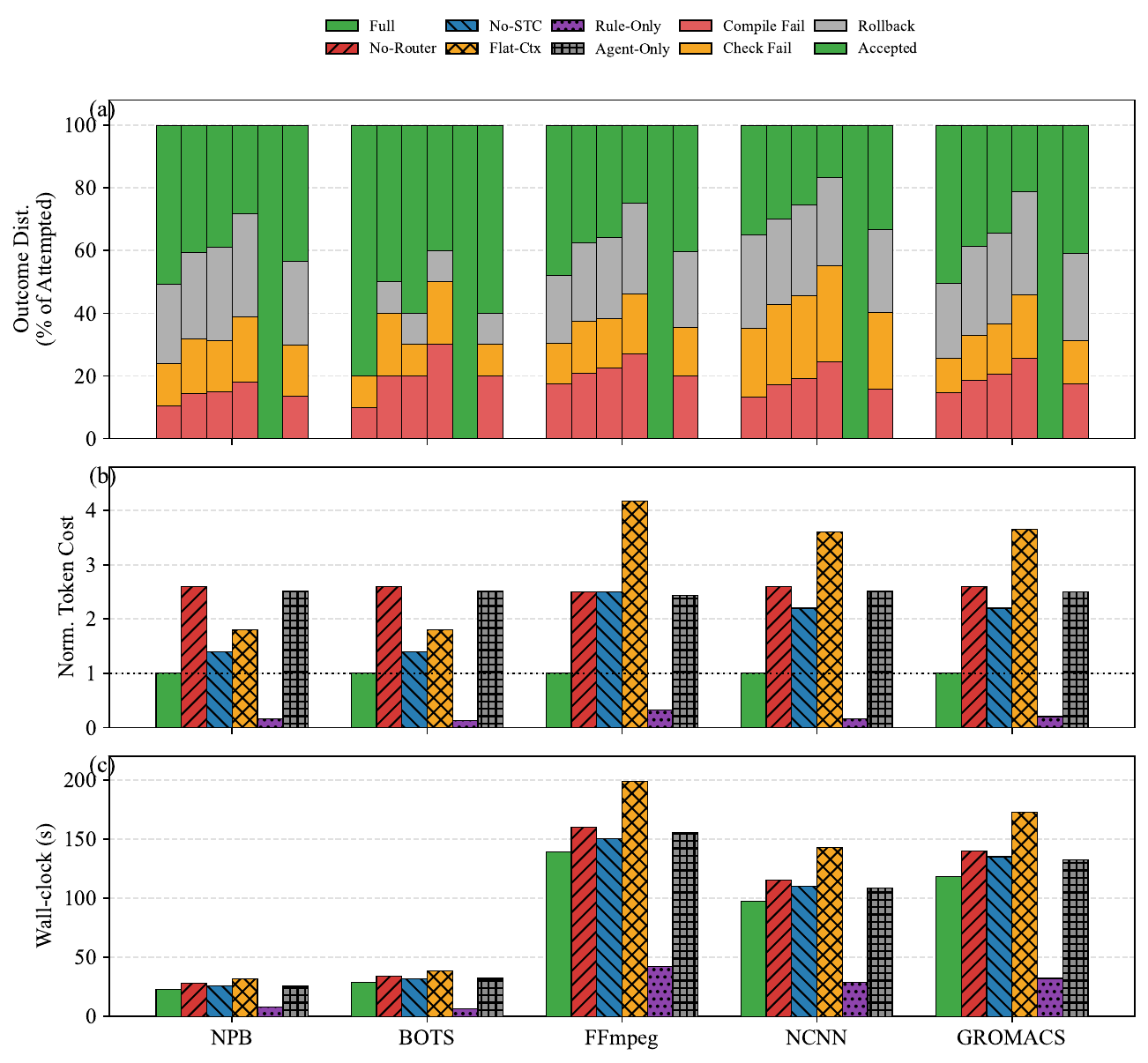}
    \vspace{0pt}
    \caption{Ablation study on RepoOMP's cost and failure behavior. Panel (a) reports the outcome distribution over compile failures, workload-check failures, rollbacks, and accepted hotspots; panel (b) reports normalized token cost; and panel (c) reports end-to-end wall-clock cost. All experiments were conducted on Intel with 16 threads, using \texttt{-O3} and Gemini 3.0 Pro.}
    \Description{A three-panel bar chart comparing the full RepoOMP pipeline against No-Router, No-STC, Flat-Ctx, Rule-Only, and Agent-Only across NPB, BOTS, FFmpeg, NCNN, and GROMACS. The top panel shows compile-failure, check-failure, rollback, and accepted-outcome shares; the middle panel shows normalized token cost; and the bottom panel shows wall-clock cost. The full system is cheapest among the agent-using variants, while Flat-Ctx is the most expensive context ablation and Rule-Only is cheapest but accepts far fewer hotspots.}
    \label{fig:ablation_token_cost}
  \end{minipage}
\end{figure}

We first ask whether the integrated design remains stronger than simpler variants built from the same overall workflow. Figure~\ref{fig:ablation_speedup} reports accepted-hotspot rate and accepted-set mean speedup, while Figure~\ref{fig:ablation_token_cost} reports outcome distribution, normalized token cost, and end-to-end wall-clock cost. Table~\ref{tab:ablation_funnel} complements these aggregate plots with hotspot-level funnel statistics for the three measured branch-removal/context-flattening variants: Flat-Ctx, Agent-Only, and Rule-Only.

\begin{table*}[t]
\centering
\caption{Hotspot-level ablation funnel and end-to-end cost for the measured component variants. \textit{Hotspots} denotes profiled candidate hotspots. \textit{Route(H/M/L)} gives the high-, medium-, and low-confidence routing counts under the reported variant. \textit{Attempted} counts candidates for which the variant attempted a source transformation. \textit{Outcome(CF/Chk/RB/Acc)} denotes compilation failures, workload-check failures, no-speedup rollbacks, and accepted hotspots. \textit{Acc.(P/A)\%} reports acceptance over profiled hotspots and over attempted transformations; \textit{Total(s)} and \textit{Token(k)} are average end-to-end wall-clock cost and model-side token usage.}
\label{tab:ablation_funnel}
\scriptsize
\renewcommand{\arraystretch}{1.06}
\setlength{\tabcolsep}{2.5pt}
\resizebox{\textwidth}{!}{%
\begin{tabular}{ccccccccc}
\toprule
Workload & Variant & Hotspots & Route(H/M/L) & Attempted & Outcome(CF/Chk/RB/Acc) & Acc.(P/A)\% & Total(s) & Token(k) \\
\midrule
\multirow{4}{*}{NPB}
 & Full & 78 & 5/62/11 & 67 & 7/9/17/34 & 43.6/50.7 & 23.1 & 76.7 \\
 & Flat-Ctx & 78 & 5/62/11 & 67 & 12/14/22/19 & 24.4/28.4 & 31.8 & 138.2 \\
 & Agent-Only & 78 & 5/62/11 & 67 & 9/11/18/29 & 37.2/43.3 & 25.4 & 192.5 \\
 & Rule-Only & 78 & 5/62/11 & 5 & 0/0/0/5 & 6.4/100.0 & 8.2 & 12.1 \\
\addlinespace
\multirow{4}{*}{BOTS}
 & Full & 11 & 1/9/1 & 10 & 1/1/0/8 & 72.7/80.0 & 28.6 & 65.9 \\
 & Flat-Ctx & 11 & 1/9/1 & 10 & 3/2/1/4 & 36.4/40.0 & 38.4 & 118.7 \\
 & Agent-Only & 11 & 1/9/1 & 10 & 2/1/1/6 & 54.5/60.0 & 32.1 & 165.3 \\
 & Rule-Only & 11 & 1/9/1 & 1 & 0/0/0/1 & 9.1/100.0 & 6.5 & 8.4 \\
\addlinespace
\multirow{4}{*}{FFmpeg}
 & Full & 552 & 33/428/91 & 461 & 80/60/100/221 & 40.0/47.9 & 139.2 & 58.7 \\
 & Flat-Ctx & 552 & 33/428/91 & 461 & 125/88/133/115 & 20.8/24.9 & 198.5 & 245.3 \\
 & Agent-Only & 552 & 33/428/91 & 461 & 92/72/111/186 & 33.7/40.3 & 155.7 & 142.8 \\
 & Rule-Only & 552 & 33/428/91 & 33 & 0/0/0/33 & 6.0/100.0 & 42.3 & 18.6 \\
\addlinespace
\multirow{4}{*}{NCNN}
 & Full & 147 & 11/103/33 & 114 & 15/25/34/40 & 27.2/35.1 & 97.5 & 88.5 \\
 & Flat-Ctx & 147 & 11/103/33 & 114 & 28/35/32/19 & 12.9/16.7 & 142.6 & 318.4 \\
 & Agent-Only & 147 & 11/103/33 & 114 & 18/28/30/38 & 25.9/33.3 & 108.3 & 221.7 \\
 & Rule-Only & 147 & 11/103/33 & 11 & 0/0/0/11 & 7.5/100.0 & 28.7 & 14.2 \\
\addlinespace
\multirow{4}{*}{GROMACS}
 & Full & 163 & 10/127/26 & 137 & 20/15/33/69 & 42.3/50.4 & 118.4 & 29.0 \\
 & Flat-Ctx & 163 & 10/127/26 & 137 & 35/28/45/29 & 17.8/21.2 & 172.3 & 105.8 \\
 & Agent-Only & 163 & 10/127/26 & 137 & 24/19/38/56 & 34.4/40.9 & 132.6 & 72.5 \\
 & Rule-Only & 163 & 10/127/26 & 10 & 0/0/0/10 & 6.1/100.0 & 32.1 & 5.8 \\
\bottomrule
\end{tabular}
}
\end{table*}

The first pattern is that context quality matters more than merely preserving an agent in the loop. Removing the bounded STC (No-STC) already lowers both accepted-hotspot rate and accepted-set mean speedup relative to the full system, and replacing multi-granularity repository evidence with a flatter function-level view (Flat-Ctx) degrades the system further. Table~\ref{tab:ablation_funnel} quantifies this effect for the measured Flat-Ctx variant: accepted hotspots drop from 34 to 19 on NPB, from 221 to 115 on FFmpeg, from 40 to 19 on NCNN, and from 69 to 29 on GROMACS, while token cost rises from 76.7k to 138.2k on NPB, from 58.7k to 245.3k on FFmpeg, from 88.5k to 318.4k on NCNN, and from 29.0k to 105.8k on GROMACS. Figure~\ref{fig:ablation_token_cost} shows the same pattern at the aggregate level: Flat-Ctx is the most expensive among the context-construction ablations in both normalized token cost and wall-clock time. We interpret this result as evidence consistent with the mechanism that repository-scale optimization benefits from preserving structured dependency evidence, not merely from exposing more raw code.

The second pattern is that routing is associated with better efficiency and fewer observed failure outcomes. In Figure~\ref{fig:ablation_speedup}, No-Router and Agent-Only remain much closer to the full system than Rule-Only on accepted-set mean speedup, indicating that the agent can recover many opportunities once enough evidence is present. However, Figure~\ref{fig:ablation_token_cost} shows that No-Router and Agent-Only pay a clear cost penalty: they increase failure mass and substantially raise both normalized token cost and wall-clock time across all five workloads. The measured Agent-Only rows in Table~\ref{tab:ablation_funnel} make this concrete. Relative to the full system, accepted hotspots fall only moderately on FFmpeg (221 to 186), NCNN (40 to 32), and GROMACS (69 to 57), but token cost rises from 58.7k to 142.8k on FFmpeg, from 88.5k to 221.7k on NCNN, and from 29.0k to 72.5k on GROMACS. These observations support the interpretation that the router is a practical control layer that limits unnecessary agent usage and keeps retry costs bounded.

Finally, removing the LLM branch sharply collapses coverage even though it produces the cheapest pipeline. Rule-Only accepts far fewer hotspots on every workload and leaves much of the hotspot funnel untouched: attempted transformations drop from 67 to 5 on NPB, from 461 to 33 on FFmpeg, from 114 to 11 on NCNN, and from 137 to 10 on GROMACS. Accepted hotspots fall accordingly, for example from 221 to 33 on FFmpeg and from 69 to 10 on GROMACS. Figure~\ref{fig:ablation_speedup} shows the same trend in accepted-hotspot rate and accepted-set mean speedup, while Figure~\ref{fig:ablation_token_cost} shows why the variant is superficially cheap: it avoids most of the expensive agent-side search by conservatively refusing the semantically harder cases. The result identifies a task-interface gap in the evaluated protocol rather than a universal limitation of traditional tools: given a command-level workload, a large profiled repository, and many candidate files, a purely local rule path lacks the evidence-recovery and search mechanism needed to decide which candidates to optimize and how to do so without breaking the executable checks. The search space is the product of profiled files, candidate loops, possible dependency contexts, and workload-specific validation outcomes; without MAP-style pruning, a conservative local tool has little basis for choosing safe non-local transformations and therefore tends to skip them. The contrast is consistent with the intended division of labor in RepoOMP: rules provide a low-cost path for structurally clear cases, whereas the agent recovers the broader set of hotspots that require richer repository-context reasoning.

\subsubsection{\textbf{Impact of Thread Count}}

To study scalability, we vary the number of threads from 16 to 32 and 64 using Gemini 3.0 Pro. The main pattern is that additional threads help when parallel work remains available, but the marginal benefit decreases once memory bandwidth, synchronization, or limited task granularity becomes dominant.

\begin{figure*}[t]
  \centering
  \begin{minipage}[t]{0.48\textwidth}
    \centering
    \includegraphics[width=0.90\linewidth]{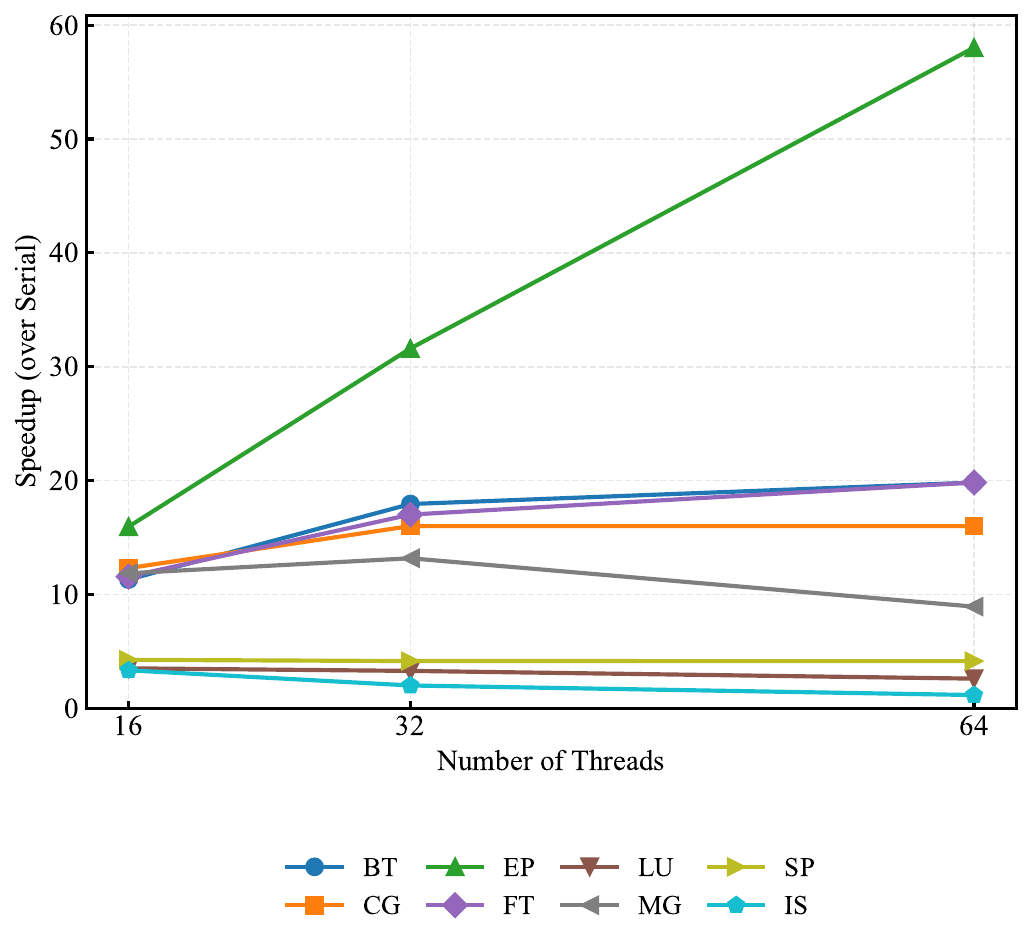}
    \vspace{0pt}
    \caption{NPB speedup with varying thread counts on Intel using -O3 and RepoOMP by Gemini 3.0 Pro.}
    \Description{A line chart showing NPB speedup as the thread count increases from 16 to 32 and 64. Some kernels continue scaling well, while others plateau or slightly decline at higher thread counts, indicating limits from bandwidth, synchronization, or available parallel work.}
    \label{fig:npb_speedup}
  \end{minipage}\hfill
  \begin{minipage}[t]{0.48\textwidth}
    \centering
    \includegraphics[width=0.90\linewidth]{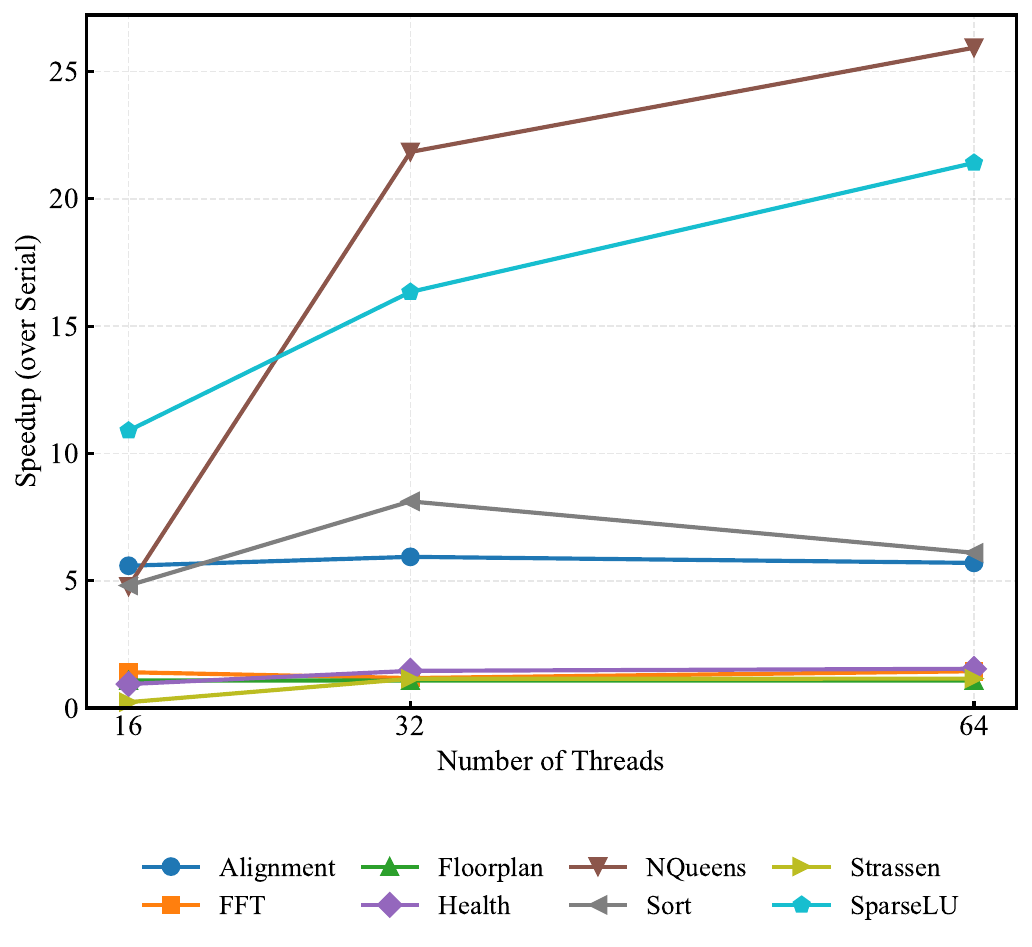}
    \vspace{0pt}
    \caption{BOTS speedup with varying thread counts on Intel using -O3 and RepoOMP by Gemini 3.0 Pro.}
    \Description{A line chart showing BOTS speedup across increasing thread counts. The curves indicate that several workloads benefit from more threads but eventually show diminishing returns, reflecting workload-dependent scalability under task-parallel execution.}
    \label{fig:bots_speedup}
  \end{minipage}
\end{figure*}

For the benchmark suites, the scaling curves mainly reveal the limits of available parallelism. As shown in Figures~\ref{fig:npb_speedup} and \ref{fig:bots_speedup}, kernels such as EP and SparseLU already obtain substantial speedup at 16 threads and continue to benefit at 32 threads. At 64 threads, however, several workloads plateau or show diminishing returns. For example, MG drops from about $13.2\times$ to $9.0\times$. This behavior is consistent with RepoOMP exposing useful parallel structure, with the remaining bottlenecks increasingly determined by workload characteristics rather than by obviously missed parallel opportunities.

\begin{figure*}[t]
  \centering
  \begin{minipage}[t]{0.48\textwidth}
    \centering
    \includegraphics[width=0.90\linewidth]{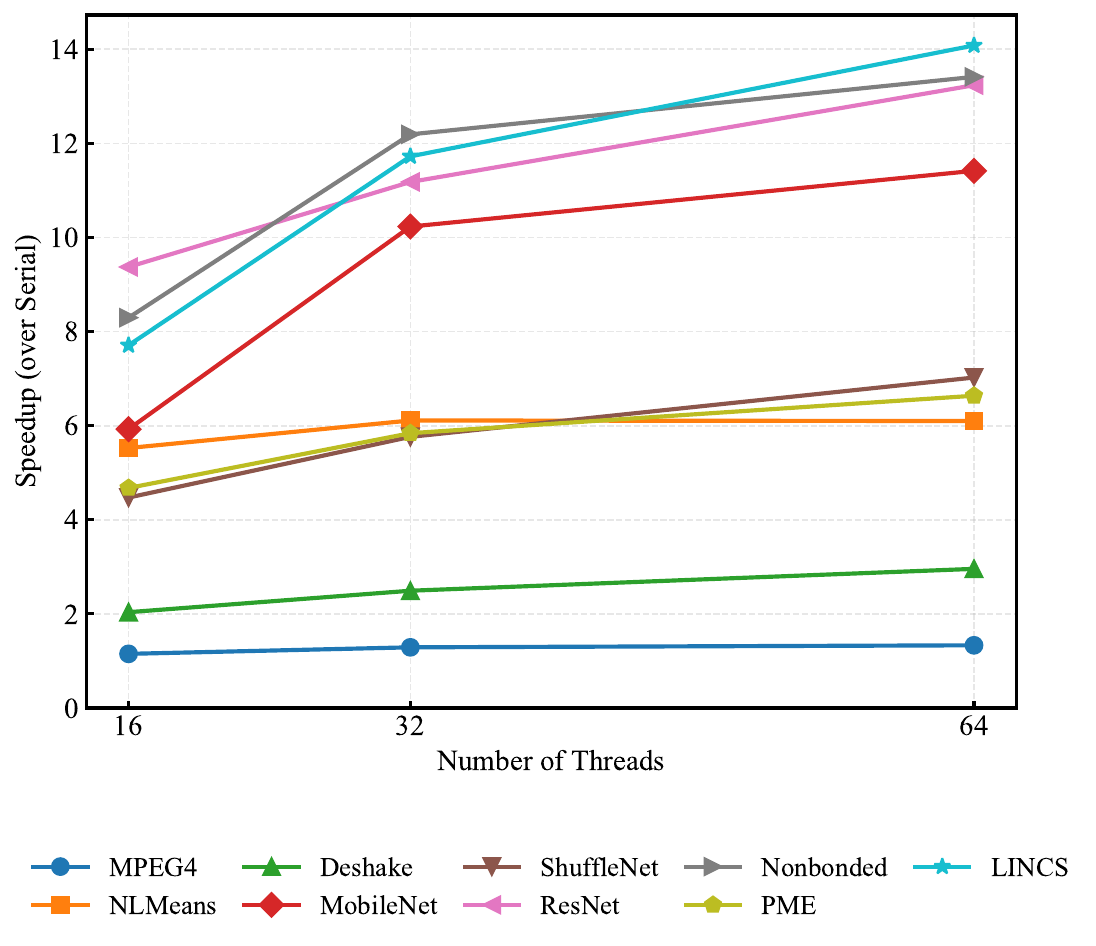}
    \vspace{0pt}
    \caption{Real-world application speedup with varying thread counts on Intel using -O3 and RepoOMP by Gemini 3.0 Pro.}
    \Description{A line chart showing the speedup of several real-world application kernels as the thread count increases. Unlike some benchmark kernels, the real-world workloads generally continue to improve with more threads, indicating that RepoOMP can expose scalable parallel regions in larger repositories.}
    \label{fig:real_world_speedup}
  \end{minipage}\hfill
  \begin{minipage}[t]{0.48\textwidth}
    \centering
    \includegraphics[width=0.90\linewidth]{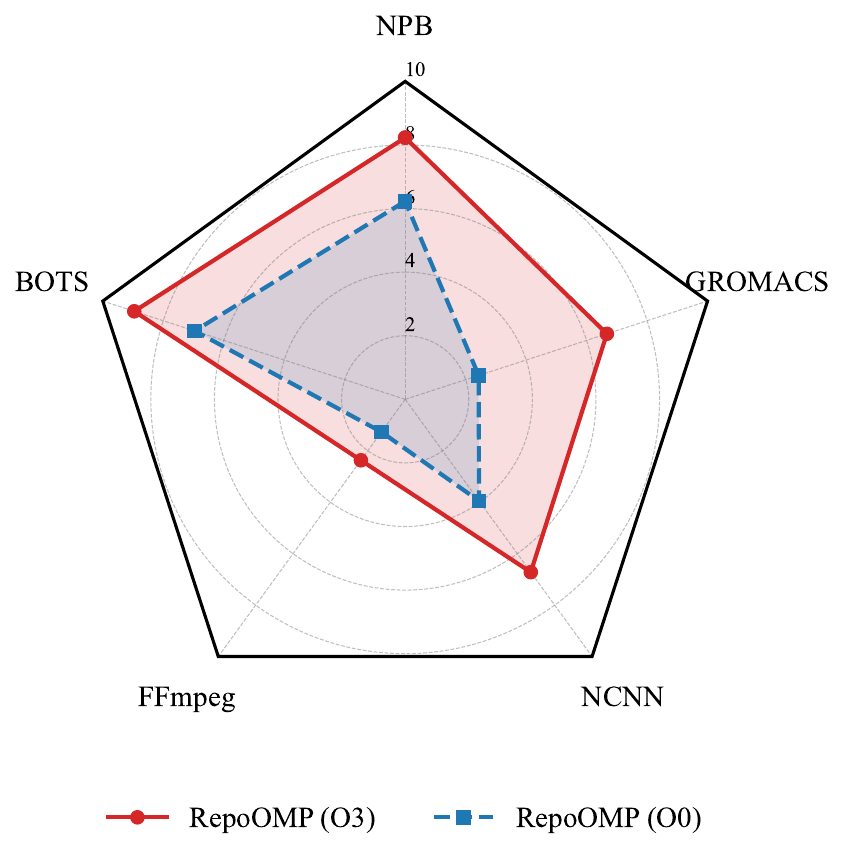}
    \vspace{0pt}
    \caption{RepoOMP speedup under -O0 and -O3 on Intel with 16 threads using Gemini 3.0 Pro.}
    \Description{A radar chart comparing RepoOMP speedup under two compiler settings, -O0 and -O3, across multiple workloads. The chart shows that compiler optimization and source-level parallelization are complementary, with the largest gaps appearing on more complex real-world applications.}
    \label{fig:Compiler}
  \end{minipage}
\end{figure*}

The real-world applications show a different trend. Figure~\ref{fig:real_world_speedup} indicates that workloads such as ResNet, LINCS, and MobileNet continue to improve as the thread count increases to 64. For instance, LINCS rises from $7.7\times$ at 16 threads to $14.1\times$ at 64 threads, and ResNet increases from $9.4\times$ to $13.3\times$. These results suggest, within the reported workloads, that RepoOMP can identify parallel regions whose granularity remains useful on many-core systems, which is particularly important for larger repositories with sustained compute intensity.

\subsubsection{\textbf{Impact of Compiler Optimization}}

Parallelism extraction and compiler optimization make complementary contributions. Figure~\ref{fig:Compiler} compares RepoOMP under \texttt{-O0} and \texttt{-O3} to evaluate how source-level OpenMP insertion interacts with backend optimization.

The effect is modest on the benchmark suites. For NPB and BOTS, the gap between \texttt{-O0} and \texttt{-O3} remains relatively small. For example, BOTS improves from about $7.5\times$ without optimization to $8.6\times$ under \texttt{-O3}, while NPB increases from $6.2\times$ to $8.3\times$. These workloads obtain much of their benefit from exposing thread-level parallelism, so compiler optimization mainly refines an already effective decomposition.

The real-world applications show a larger gap. GROMACS increases from $3.2\times$ at \texttt{-O0} to $7.2\times$ at \texttt{-O3}, and NCNN improves from $5.5\times$ to nearly $9.0\times$. This pattern suggests that RepoOMP provides the parallel structure, while compiler optimizations such as vectorization and loop-level refinement help realize more of the available speedup in complex repositories.

\subsubsection{\textbf{Impact of CPU Architecture}}

Parallelization strategies often respond differently to hardware topology, cache behavior, and memory hierarchy. We therefore compare RepoOMP and the agent baselines on both Intel and AMD CPUs using the benchmark suites and the real-world applications.

\begin{figure*}[t]
  \centering
  \begin{minipage}[t]{0.48\textwidth}
    \centering
    \includegraphics[width=0.91\linewidth]{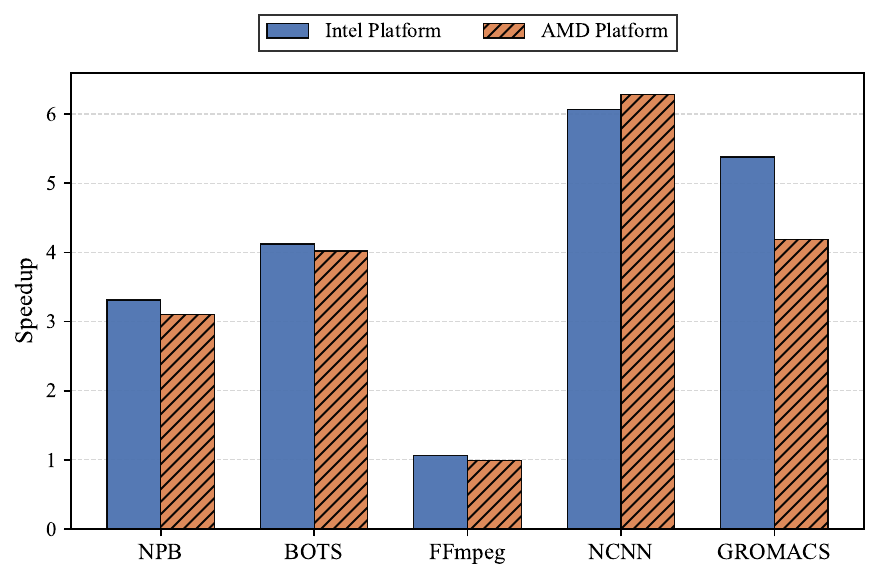}
    \vspace{0pt}
    \caption{Claude Code speedup across CPU architectures using 16 threads and -O3.}
    \Description{A grouped bar or radar-style comparison of Claude Code speedup on Intel and AMD platforms across several workloads. The figure highlights that Claude Code can produce useful optimizations but shows noticeable sensitivity to architecture-specific differences on some tasks.}
    \label{fig:cpu_cc}
  \end{minipage}\hfill
  \begin{minipage}[t]{0.48\textwidth}
    \centering
    \includegraphics[width=0.91\linewidth]{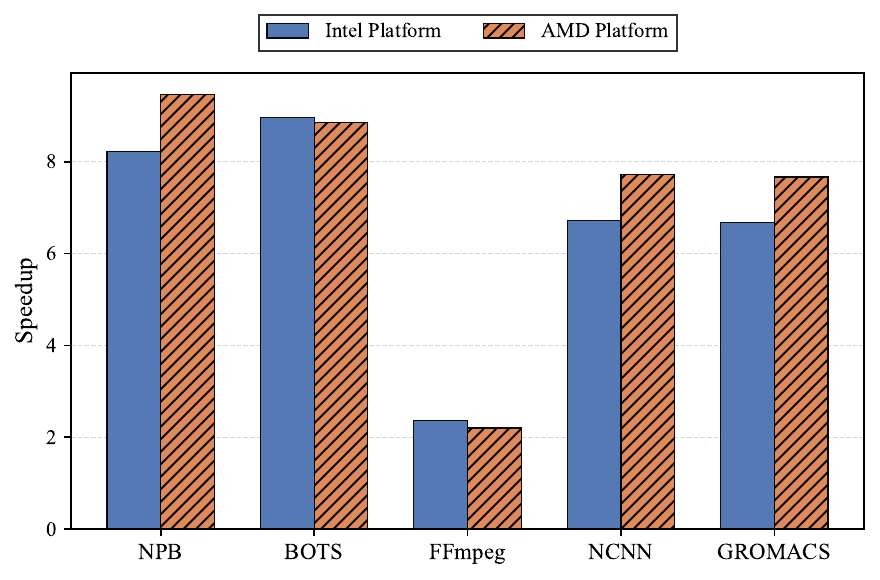}
    \vspace{0pt}
    \caption{RepoOMP speedup across CPU architectures using 16 threads and -O3.}
    \Description{A grouped bar or radar-style comparison of RepoOMP speedup on Intel and AMD platforms across several workloads. The figure shows that RepoOMP sustains strong performance on both measured architectures and in some cases achieves higher gains on AMD.}
    \label{fig:cpu_RepoOMP}
  \end{minipage}
\end{figure*}

Figure~\ref{fig:cpu_cc} shows that the Claude Code baseline is more variable across the two measured architectures on some workloads, including GROMACS. These differences suggest architecture-specific variation in the reported two-platform study.

In the reported two-platform study, RepoOMP shows more stable behavior across the two platforms. Figure~\ref{fig:cpu_RepoOMP} indicates that the framework maintains strong performance on Intel and, on several workloads, performs even better on AMD. For example, RepoOMP reaches $9.4\times$ on AMD versus $8.2\times$ on Intel for NPB, and $7.8\times$ versus $6.8\times$ for NCNN. One plausible interpretation is that grounding the optimization process in explicit dependency and hotspot evidence reduces reliance on loosely matched repository context; the current evidence, however, does not isolate architecture effects from workload and implementation choices.

\section{Discussion}

RepoOMP is most suitable for repositories in which performance-critical regions can be localized and the relevant dependency structure can be summarized with sufficient fidelity, such as dense numerical kernels, multimedia pipelines, and molecular dynamics components. In these settings, MAP isolates the code and shared-state evidence needed for transformation, while the router avoids invoking the agent on cases that are already clear or evidently unsafe.

Several classes of programs remain difficult for the current design. Heavy pointer aliasing, hard-to-summarize dynamic memory behavior, and I/O-dominated paths tend to produce conservative dependency summaries. Algorithms whose shared-state updates require substantial refactoring are also outside the present scope of automated OpenMP insertion.

RepoOMP also incurs nontrivial preprocessing cost. Building MAP requires repository-level static analysis, and the profiling phase depends on representative executions and standard profiling tools such as \texttt{gprof}. The framework further assumes that the target repository can be compiled successfully before optimization begins. Missing dependencies or unstable build configurations can therefore limit practical applicability.

Several limitations remain. The real-world evidence discloses workflow denominators, repository-level distributions, and nine detailed accepted kernels, while per-candidate blocker traces and retry histories are summarized at the workflow level rather than expanded for every accepted hotspot. Token-cost analysis captures model-side usage, and Table~\ref{tab:candidate_selection} reports the corresponding end-to-end wall-clock cost. For the nine real-world case studies, we report five-run 16-thread means together with 1/16/32/64-thread executable checks, which gives a compact view of stability across the reported workloads and settings.

\textbf{What the evaluation does not establish.} The reported evidence should be read with several boundaries. First, acceptance means compilation, workload-specific executable checks, and positive speedup under the tested inputs; it is not a proof of race freedom, schedule-independent correctness, or exhaustive semantic preservation. Second, the 330 accepted real-world hotspots support repository-scale coverage and accepted-set speedup distributions, whereas matched-backbone speedup, token-cost, repeated-run, ThreadSanitizer, and cross-thread evidence are reported for the nine detailed real-world kernels used for robustness and cost analyses. Third, the results do not establish universal superiority over all automatic parallelizers or all possible repository-agent baselines. The traditional-tool comparison mainly characterizes the command-level hotspot-discovery protocol used in this study, where local rule tools lack a built-in mechanism for moving from an executable workload command to the relevant optimization files and safe transformations across a large repository; it is not an oracle-targeted loop-transformation comparison. Finally, the Intel/AMD study is a two-platform sensitivity check rather than a broad architecture-generalization claim.

Because MAP represents dependency and hotspot evidence rather than OpenMP-specific syntax alone, the workflow could be adapted to other optimization targets such as MPI or CUDA by changing the downstream transformation policy while retaining repository-level evidence recovery and verification.

\begin{table*}[!t]
  \centering
  \caption{Repository-level speedup distribution across all 330 accepted real-world hotspots. Percentiles and bins are computed over accepted hotspots only; the `Total real-world' row aggregates FFmpeg, NCNN, and GROMACS.}
  \label{tab:accepted_distribution_full}
  \scriptsize
  \renewcommand{\arraystretch}{1.05}
  \setlength{\tabcolsep}{2.2pt}
  \resizebox{\textwidth}{!}{%
  \begin{tabular}{ccccccccccccc}
    \toprule
    Repo & Accepted & P10 & P25 & Median & P75 & P90 & Min & Max & Below 1.2$\times$ & 1.2--1.5$\times$ & 1.5--2.0$\times$ & At least 2.0$\times$ \\
    \midrule
    FFmpeg & 221 & 1.15x & 1.35x & 1.85x & 3.20x & 5.50x & 1.05x & 12.40x & 30 & 60 & 50 & 81 \\
    NCNN & 40 & 1.40x & 2.10x & 4.50x & 6.80x & 9.50x & 1.10x & 15.20x & 2 & 5 & 8 & 25 \\
    GROMACS & 69 & 1.30x & 1.80x & 3.60x & 6.20x & 8.50x & 1.08x & 14.80x & 5 & 10 & 15 & 39 \\
    \midrule
    Total real-world & 330 & 1.18x & 1.45x & 2.25x & 4.80x & 7.50x & 1.05x & 15.20x & 37 & 75 & 73 & 145 \\
    \bottomrule
  \end{tabular}
  }
\end{table*}

\subsection{Threats to Validity}

\textbf{Construct and statistical conclusion validity.} Our acceptance criterion is operational: a candidate must compile, pass the workload-specific executable checks used in the evaluation, and yield measurable speedup on the tested workload. For the nine detailed real-world kernels, we report five-run means at 16 threads, disclose per-kernel spread, exercise the checks across 1/16/32/64 threads, and add a matched dynamic-race audit in Table~\ref{tab:tsan_summary}. These steps strengthen the empirical picture for the reported workloads, but they do not establish schedule-independent correctness, race freedom, or exhaustive statistical coverage beyond the tested inputs and configurations.

\textbf{Internal validity.} RepoOMP contains heuristic decisions, especially in routing and context construction. We expose those design choices through ablations and repository-level outcome counts, which supports the usefulness of the workflow architecture as a system. At the same time, the current evidence does not prove that every routing decision or heuristic threshold is individually optimal for every hotspot.

\textbf{External validity and reproducibility.} The evaluation combines benchmark suites with a repository-scale workflow audit over three large real-world codebases and detailed matched-backbone results on nine accepted case-study kernels. This supports the claim that RepoOMP can operate effectively on repository-scale software with substantial compute kernels, but it does not yet establish coverage for the most alias-heavy, I/O-bound, or heavily refactored paths.

\begin{table}[t]
  \centering
  \caption{Within-repository positions of the nine detailed real-world case studies. \textit{Spd.16T} is the five-run mean 16-thread speedup, \textit{Rank} is the kernel's speedup rank inside the repository's accepted set, \textit{Accepted} is the size of that accepted set, and \textit{Rank\%} is Rank divided by Accepted. Lower rank and lower rank percentile indicate a faster position inside the repository's accepted set at 16 threads.}
  \label{tab:accepted_case_rank}
  \footnotesize
  \renewcommand{\arraystretch}{1.03}
  \setlength{\tabcolsep}{4pt}
  \begin{tabular}{cccccc}
    \toprule
    Repo & Kernel & Spd.16T & Rank & Accepted & Rank\% \\
    \midrule
    FFmpeg & MPEG4 & 1.13x & 195 & 221 & 88\% \\
    FFmpeg & NLMeans & 5.29x & 25 & 221 & 11\% \\
    FFmpeg & Deshake & 1.94x & 105 & 221 & 47\% \\
    NCNN & MobileNet & 5.65x & 12 & 40 & 30\% \\
    NCNN & ShuffleNet & 5.02x & 16 & 40 & 40\% \\
    NCNN & ResNet & 9.91x & 3 & 40 & 7\% \\
    GROMACS & Nonbd & 8.23x & 7 & 69 & 10\% \\
    GROMACS & PME & 4.52x & 28 & 69 & 40\% \\
    GROMACS & LINCS & 6.40x & 15 & 69 & 21\% \\
    \bottomrule
  \end{tabular}
\end{table}

\section{Conclusion and Future Work}

Hotspot-oriented OpenMP auto-parallelization inside large repositories remains difficult because the evidence needed to judge loop safety and profitability is often distributed across files, helper functions, and shared state. Rule-based tools lose recall when that evidence is no longer locally provable, whereas agent-based systems become unreliable when the prompt omits a critical dependency or includes too much unrelated repository context.

RepoOMP addresses this problem by recovering repository-level evidence before generation. MAP captures dependency and hotspot information across repository, file, and function scopes, and the Rule-Agent Router uses that evidence to decide whether a candidate should be handled by deterministic rules, by the agent, or by a conservative low-confidence path. This design allows the agent to operate on a bounded context that contains the facts required for transformation instead of a large, noisy repository slice.

Our evaluation combines a hotspot-level workflow audit over 951 profiled hotspots with detailed task-level results on nine accepted real-world kernels drawn from a larger accepted real-world set of 330 repository hotspots. Among that broader accepted set, the disclosed distribution has median $2.25\times$, interquartile range $1.45$--$4.80\times$, and 145 cases at least $2.0\times$, while the nine detailed kernels span repository-internal ranks from $3/40$ to $195/221$. On the reported tasks, RepoOMP achieves average speedups of $8.23\times$ and $8.96\times$ on NPB and BOTS, respectively; for the nine detailed accepted real-world kernels used for matched-backbone and robustness analyses, it achieves $5.25\times$ while substantially reducing agent-side token cost relative to the unstructured agent baselines used in this study. These results support RepoOMP as an evidence-guided workflow for command-level hotspot parallelization within the evaluated repository settings.

Future work will focus on improving dependency precision for difficult cases such as alias-heavy code, broadening support for transformations that require deeper restructuring, and extending the framework to additional optimization targets and hardware settings.


\bibliographystyle{unsrtnat}
\bibliography{references}

\end{document}